*Research Article*

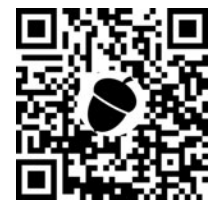

Open camera or QR reader and
scan code to access this article
and other resources online.

# Protective Effects of Halite
to Vacuum and Vacuum-Ultraviolet Radiation:
A Potential Scenario During a Young Sun Superflare

Ximena C. Abrevaya,[1,2,*,†] Douglas Galante,[3] Paula M. Tribelli,[4,5] Oscar J. Oppezzo,[6,†] Felipe Nóbrega,[7]
Gabriel G. Araujo,[8] Fabio Rodrigues,[9] Petra Odert,[10,†] Martin Leitzinger,[10,†] Martiniano M. Ricardi,[11]
Maria Eugenia Varela,[12] Tamires Gallo,[13] Jorge Sanz-Forcada,[14] Ignasi Ribas,[15,16] Gustavo F. Porto de Mello,[17,†]
Florian Rodler,[18] Maria Fernanda Cerini,[3] Arnold Hanslmeier,[10,†] and Jorge E. Horvath[19,†]

## Abstract

Halite (NaCl mineral) has exhibited the potential to preserve microorganisms for millions of years on Earth.
This mineral was also identified on Mars and in meteorites. In this study, we investigated the potential of halite
crystals to protect microbial life-forms on the surface of an airless body (*e.g.*, meteorite), for instance, during a
lithopanspermia process (interplanetary travel step) in the early Solar System. To investigate the effect of the
radiation of the young Sun on microorganisms, we performed extensive simulation experiments by employing a
synchrotron facility. We focused on two exposure conditions: vacuum (low Earth orbit, $10^{-4}$ Pa) and vacuum-
ultraviolet (VUV) radiation (range 57.6–124 nm, flux 7.14 W/m²), with the latter representing an extreme
scenario with high VUV fluxes comparable to the amount of radiation of a stellar superflare from the young
Sun. The stellar VUV parameters were estimated by using the very well-studied solar analog of the young Sun,
$\kappa^1$ Cet. To evaluate the protective effects of halite, we entrapped a halophilic archaeon (*Haloferax volcanii*) and
a non-halophilic bacterium (*Deinococcus radiodurans*) in laboratory-grown halite. Control groups were cells entrapped in salt crystals (mixtures of different salts and NaCl) and non-trapped (naked) cells, respectively. All groups were exposed either to vacuum alone or to vacuum plus VUV. Our results demonstrate that halite can serve as protection against vacuum and VUV radiation, regardless of the type of microorganism. In addition, we found that the protection is higher than provided by crystals obtained from mixtures of salts. This extends the protective effects of halite documented in previous studies and reinforces the possibility to consider the crystals of this mineral as potential preservation structures in airless bodies or as vehicles for the interplanetary transfer of microorganisms. Key Words: Lithopanspermia—Microorganisms—Origin of life—Meteorites—Astrobiology—VUV. Astrobiology 23, xxx–xxx.

[1]Instituto de Astronomía y Física del Espacio (IAFE), UBA–CONICET, Pabellón IAFE, Ciudad Universitaria, Ciudad Autónoma de
Buenos Aires, Argentina.
[2]Facultad de Ciencias Exactas y Naturales, Universidad de Buenos Aires, Ciudad Universitaria, Ciudad Autónoma de Buenos Aires,
Argentina.
[3]CNPEM—LNLS, Campinas, Brazil.
[4]Departamento de Química Biológica, Facultad de Ciencias Exactas y Naturales, Universidad de Buenos Aires, Ciudad Autónoma de
Buenos Aires, Argentina.
[5]IQUIBICEN, CONICET, Universidad de Buenos Aires, Ciudad Autónoma de Buenos Aires, Argentina.
[6]Comisión Nacional de Energía Atómica, Buenos Aires, Centro Atómico Constituyentes, Argentina.
[7]University of São Paulo, São Paulo, Brazil.
[8]Department of Microbiology, Institute of Biomedical Sciences, University of São Paulo, São Paulo, Brazil.
[9]Instituto de Química, Universidade de São Paulo, São Paulo, Brazil.
[10]Institute of Physics, University of Graz, Graz, Austria.
[11]Instituto de Fisiología, Biología Molecular y Neurociencias (IFIByNE-CONICET), Departamento de Fisiología y Biología Molecular y
Celular (FBMC), Facultad de Ciencias Exactas y Naturales, Universidad de Buenos Aires, Ciudad Autónoma de Buenos Aires, Argentina.
[12]Synchrotron Radiation Research, Lund University, Lund, Sweden.
[13]Instituto de Ciencias Astronómicas de la Tierra y del Espacio (ICATE-CONICET), San Juan, Argentina.
[14]Centro de Astrobiología (CSIC-INTA), ESAC Campus, Madrid, Spain.
[15]Institut de Ciències de l'Espai (ICE, CSIC), Campus UAB, Bellaterra, Spain.
[16]Institut d'Estudis Espacials de Catalunya (IEEC), Barcelona, Spain.
[17]Observatorio do Valongo, Universidade Federal do Rio de Janeiro, Rio de Janeiro, Brazil.
[18]European Southern Observatory, Santiago de Chile, Chile.
[19]Instituto de Astronomía, Geofísica e Ciencias Atmosfericas (IAG), Universidade de Sao Paulo, Sao Paulo, Brazil.
*Invited Scientist: Institute of Physics, University of Graz, Graz, Austria.
†Members of the BioSun project.

## 1. Introduction

**F**OR MANY DECADES, diverse astrobiological studies have been carried out to determine the underlying biological mechanisms and chances of survival of microorganisms exposed to the outer space environment. These include determination of the effects of exposure to extreme conditions present in the interplanetary medium, featuring radiation, vacuum, very low temperatures, and microgravity, among others, alone or in various combinations.

Part of these experiments took place in outer space (for instance, the International Space Station, on rocket flights/satellites) and include the Biopan, PERSEUS, EXPOSE missions, and more recently the ''Tanpopo'' mission (*e.g.*, Demets *et al.*, 2005; Kawaguchi *et al*., 2016; Mancinelli *et al*., 1998; Rabbow *et al.*, 2012; Yamagishi *et al.*, 2018; see also the reviews by Horneck, 2010; Moissl-Eichinger *et al.*, 2016; Olsson-Francis and Cockell, 2010; and references therein).

Space experiments were complemented by laboratory experimental simulations on the ground (see, *e.g.*, Horneck *et al.*, 2010; Moissl-Eichinger *et al.*, 2016; Olsson-Francis and Cockell, 2010). Some of these simulation experiments have made use of synchrotron accelerators with the goal to reproduce certain radiation wavelength regimes that are characteristic of extraterrestrial environments, such as vacuum-ultraviolet (VUV) radiation (*e.g.*, Abrevaya *et al.*, 2011; Araujo *et al.*, 2020; Paulino-Lima *et al.*, 2010).

These efforts, in general, involve the feasibility of natural processes of interplanetary transfer of microbial life, widely known as the Panspermia Hypothesis (Arrhenius, 1903; Thomson, 1871; Richter, 1865) and the probability of biological contamination of other planetary bodies with terrestrial microorganisms by processes derived from human actions (for instance, by *in situ* missions with space probes or spacecrafts) known as *human-directed Panspermia* (Nicholson *et al.*, 2005), which eventually led to the development of planetary protection procedures (see *e.g.*, Moissl-Eichinger *et al.*, 2016).

The viable transfer of life from one planet to another is still a matter of debate, as it remains an open question as to whether life could endure multiple harmful conditions if exposed to the space environment. Exposure to these conditions is dramatically increased in the case of a natural Panspermia process (*e.g.*, lasting about $10^5$–$10^7$ years for transit from Mars to Earth, as estimated by Mileikowsky *et al.*, 2000) in comparison to human-directed Panspermia, which can last in the order of months or years for planetary bodies of the Solar System.

In particular, meteorites could provide protection from some of the harsh conditions of the space environment during a Panspermia process (Horneck *et al.*, 2002). This scenario of interplanetary transport of microorganisms by meteorites, that is, lithopanspermia, describes a natural process whereby microbial life inhabiting near-surface rocks could be ejected by large-scale impacts and transferred between planetary bodies (Melosh, 1988; Mileikowsky *et al.*, 2000; Stöffler *et al.*, 2007; Thomson, 1871).

This exchange of meteoritic material may have been common around 4.1–3.8 Gyr ago during the postulated Late Heavy Bombardment period (LHB), when Earth and other inner planets of the Solar System received a high influx of meteorites striking their surfaces. The LHB is very close to the date of the earliest evidence of life, based on the fossil record and carbon isotope signatures in sediments of old rocks that indicate the existence of first life-forms around 3.9–3.48 Gyr ago (Hassenkam *et al.*, 2017; Mojzsis *et al.*, 1996; Noffke *et al*., 2013; Rosing, 1999; Schidlowski, 2001; Tashiro *et al.*, 2017) or even 4.1 Gyr ago if evidence provided by the study of graphite inclusions in zircon rocks (Bell *et al.*, 2015) is considered.

In addition, Martin *et al*. (2012) supported the idea that life may even have been present during the Hadean or in the early Archean. A more recent study has ''pushed'' the epoch of the LHB back in time, thereby shortening the Hadean era and making room for a cooler early Earth even before 4 Gyr ago (Mojzsis *et al.*, 2019). The conditions on Earth during LHB are supposed to have been very extreme, and this period has been considered too harsh for life. Nevertheless, computer models used to recreate the thermal effects of LHB on the planetary lithosphere have shown that a possible terrestrial near-surface and subsurface primordial microbial biosphere could not have been completely sterilized, at least after the end of primary accretion onto the planets and the putative impact origin of the Moon (Abramov and Mojzis, 2009).

Considering all these possibilities and the proximity in time between the LHB and the appearance of the first life-forms, perhaps as early as 3.8 Gyr ago (Schidlowski, 1988, 2001), it is reasonable to speculate that, during the LHB, meteorites may have been carriers of microbial-like life-forms and transporting living matter between planetary bodies of the Solar System by natural impact processes (Nicholson, 2009).

Whether microorganisms would remain viable during the lithopanspermia process has been an unknown, given that the transferred life-forms would have to survive extreme conditions during a process divided into three main steps as follows: (1) Ejection of the material from the parent planet containing the life-forms, which implies that microorganisms would be subjected to an escape process and, therefore, would be exposed to shock pressures; (2) interplanetary travel where the microorganisms will be exposed to multiple extreme conditions in the interplanetary space environment perhaps for millions of years; and (3) entry by way of a meteorite on a recipient planet, where microorganisms would be exposed to very high temperatures due to frictional heating and hypervelocity (for a review of these steps, see Nicholson, 2009).

The interplanetary step is one of the more challenging for life, as microbial-like forms would have to survive a combination of environmental stressors from space for periods that could last millions of years. Radiation and vacuum are among the conditions considered the most deleterious for life in the interplanetary environment. Regarding radiation, the ultraviolet (UV; 200–400 nm) band is considered the component of solar radiation most immediately lethal to microorganisms, due to its high energy and efficient absorption by biological macromolecules (proteins, nucleic acids, and lipids) (Horneck et al., 2002; Nicholson et al., 2005).

For instance, experimental data have shown that extraterrestrial solar UV radiation is of the order of a thousand times biologically more efficient than UV at the surface of the Earth, and that space vacuum has a synergistic effect increasing the UV sensitivity of microorganisms (e.g., Horneck et al., 2002).

Microorganisms transported over such lengthy time scales would also have to endure, for example, starvation. However, microorganisms that have the capability to remain dormant for extended periods of time, for example, spore-forming bacteria, may be able to endure extreme conditions, and therefore, they may have an advantage to survive Panspermia (e.g., Horneck et al., 1994). There is also an accumulation of evidence that suggests that non-sporulating microorganisms may have the capacity to endure over geological timescales.

Several studies have suggested that microorganisms that inhabit hypersaline environments, such as halophilic archaea (Haloarchea), could survive while trapped inside halite (NaCl minerals) for millions of years (Norton et al., 1993; Grant et al., 1998; McGenity et al., 2000). Halite forms as brines evaporate, and microorganisms living in these environments are naturally trapped within fluid inclusions inside these crystals (e.g., Norton and Grant, 1993). Moreover, crystals of halite, in particular, and crystals of salts, in general, may provide shielding against UV radiation (Fendrihan et al., 2009; Godin et al., 2020).

Haloarchaea are, in general, considered UV-tolerant and resistant to desiccation (among other physicochemical processes) (Dornmayr-Pfaffenhuemer et al., 2011; Oren, 2014; Stan-Lotter and Fendrihan, 2015), and they can grow at relatively low oxygen concentrations due to the poor solubility of oxygen in brines, which could represent an advantage to survive inside fluid inclusions as these quickly become an anaerobic environment (McGenity et al., 2000). Even though the existence of spore-like resistant structures has not been documented for these microorganisms, some studies suggest that the cells can be dormant during these conditions when developing cyst-like structures or spheres (Grant et al., 1998; McGenity et al., 2000; Fendrihan et al., 2012).

Given the existence of analog evaporitic environments between Earth and Mars (Nisbet and Sleep, 2001; Schi-

dlowski, 2001), it could be speculated that microorganisms with similar features to haloarchaea, or the remnants of it, could still exist or have existed in martian halite (Stan-Lotter et al., 2004).

Halite crystals have also been found in different meteorite specimens (Berkley et al., 1978; Barber, 1981; Bridges and Grady, 1999; Zolensky et al., 1999). Among them is the halite found in the Nahkla meteorite (Bridges and Grady, 1999; Gooding et al., 1991), which is particularly relevant because of the meteorite's putative martian origin. The existence of hypersaline environments is not an exclusive feature of Earth and Mars; it can be extended to other planetary bodies of the Solar System, for instance, Jupiter's moon Europa (Antunes et al., 2020).

Therefore, it appears justified to consider halite as a potential structure to preserve microorganisms in an interplanetary journey. In addition, given the potential capabilities of haloarchaea to remain viable for extended periods of time inside halite fluid inclusions, these microorganisms are suitable models for evaluation of their survival in space (Mancinelli, 1998; Abrevaya et al., 2011). In addition, other microorganisms have been considered for these studies, such as Deinococcus radiodurans, a non-sporulating bacterium that has been extensively used because of its radiation and desiccation tolerance (Battista, 1997; Bauermeister et al., 2011).

The goal of the present study was to analyze the protective effects that halite crystals may confer on microorganisms entrapped inside fluid inclusions (the halophilic archaeon Haloferax volcanii, and the non-halophilic bacterium D. radiodurans) when exposed to simulated interplanetary conditions such as high vacuum ($10^{-4}$ Pa) and VUV radiation (synchrotron toroidal grating monochromator [TGM] beamline, 57.6–124 nm) at levels that could be found in the interplanetary environment.

Vacuum and VUV can reach planetary orbits and surfaces of planetary bodies that lack atmospheres, such as moons in the Solar System, or naturally occurring objects that originate in a planetary body such as meteorites or asteroids. Given that it is hypothesized that the frequency of impacts in the Solar System may have been significantly increased during the LHB, thus favoring the exchange of meteorites between planetary bodies, we chose to set the scenario of our study in a potential intersection between this period and the origin of life (3.8 Gyr ago).

Owing to its increased magnetic activity, the radiation in the wavelength regime below 170 nm (which includes the VUV) that was emitted by the young Sun during these early ages of the Solar System was much stronger than that of the present Sun (Güdel, 2007). Such enhanced radiation fluxes have particular relevance for the space environment considered in our study. As solar analogs can provide information about the characteristics of the young Sun, we used the well-investigated solar analog $\kappa^1$ Cet as a proxy.

This star has an estimated age of 0.75 Gyr, which corresponds to the Sun at 3.85 Gyr ago (coincident with the period considered in our study) and represents well the young Sun in terms of its parameters such as mass, radius, temperature, and metallicity (see, e.g., Güdel, 2007; Ribas et al., 2005, 2010). Moreover, due to its vicinity (d = 9.2 parsec [pc]) and brightness, it has been observed across the entire electromagnetic spectrum. Young stars are known to flare more frequently than older stars and additionally feature the so-called superflares (see *e.g.*, Balona, 2015; Davenport, 2016; Maehara *et al.*, 2012).

Flares can be described as sudden outbursts of radiation powered by magnetic energy, which are observable from X-rays to the radio regime and are frequently detected from both the Sun and other stars. Superflares are exceptionally strong flares with flare energies that exceed $10^{33}$ erg ($10^{26}$ J). Recent studies have revealed that young solar-like stars can have superflare frequencies of up to one per day (Doyle *et al.*, 2020; Tu *et al.*, 2020). In contrast, our present-day Sun has a much lower superflare frequency, which is estimated to be around one in a few thousand years (Shibata *et al.*, 2013).

Although the effects of exposure to detrimental conditions such as high vacuum and VUV radiation at levels that can be found in space have already been studied for different types of microorganisms (*e.g.*, Horneck, 1999; Paulino-Lima *et al.*, 2010; Abrevaya *et al.*, 2011), the survival of microorganisms subjected to VUV exposures comparable to those expected during energetic stellar events such as flares and superflares has not been investigated.

Therefore, the goal of this study, which is part of the Biosun project (Abrevaya *et al.*, 2014), is to unveil the protective effects of halite against vacuum and VUV.

## 2. Materials and Methods

### 2.1. Microorganisms and culture conditions

For the experiments in this study, two different microorganisms were used: the bacterium *D. radiodurans* (R1 wild-type strain) and the haloarchaeon *H. volcanii* (DS70). Both microorganisms were grown aerobically under orbital agitation at 200 rpm and 30°C.

*H. volcanii* was cultivated in Hv-YPC broth (Kauri *et al.*, 1990), containing (g/L): yeast extract (5); peptone (1); casaminoacids (1); NaCl (144); $MgSO_4.7H_2O$ (21); $MgCl_2.6H_2O$ (18); KCl (4.2); $CaCl_2$ (3 m*M*); and Tris-HCl (12 m*M*), with pH adjusted to 7.

*D. radiodurans* was cultivated in TGY broth (Anderson, 1956), containing tryptone (10 g/L), yeast extract (6 g/L), and glucose (2 g/L).

### 2.2. Sample preparation

Aliquots of late-exponential cultures of *D. radiodurans* were washed twice with sterilized distilled water (this has no significant effects on the loss of the viability, see: Chou and Tan, 1991). The same procedure was used with aliquots of *H. volcanii* cultures, but distilled water was replaced with saline solution (to prevent cell leakage) containing a mixture of salts (in g/L): NaCl (144); $MgSO_4.7H_2O$ (21); $MgCl_2.6H_2O$ (18); KCl (4.2); $CaCl_2$ (3 m*M*); and Tris-HCl (12 m*M*), with pH adjusted to 7 (this saline solution has the same composition of Hv-YPC broth without yeast, peptone, and casaminoacids, from now on called ''Hv-SS'').

In both cases, the $OD_{600nm}$ of the final suspension was adjusted to 1, and the final resuspension step was done in different solutions as follows, giving different cell suspensions (CS): (CS1) *D. radiodurans* cells were resuspended in distilled water; (CS2) *D. radiodurans* cells were resuspended in a NaCl 4 *M* solution (pH = 7); (CS3) *H. volcanii* cells were resuspended in a saline solution made of a salt mixture (Hv-SS) (pH = 7); and (CS4) *H. volcanii* cells were resuspended in NaCl 4 *M* saline solution (pH = 7).

Drops (1 µL) of the CS (CS1, CS2, CS3, CS4) were deposited over glass stripes and evaporated at room temperature under sterile conditions. For CS1, the evaporation procedure led to the formation of cells that were non-trapped or non-embedded (from now on called ''naked cells'') as deposited on glass slides. For CS2 and CS4, the evaporation procedure led to the formation of halite crystal-embedded cells, and for CS3, it led to the formation of salt crystal-embedded cells. Once the microorganisms were dried under different conditions, they were kept at room temperature until they were exposed to vacuum and radiation at the synchrotron facility.

### 2.3. Synchrotron parameters and TGM beamline VUV flux

The VUV emission in the interplanetary medium was simulated by using the TGM beamline at the Brazilian Synchrotron Light Laboratory (LNLS, Brazilian Center for Research in Energy and Materials [CNPEM], Campinas, Brazil).

We employed a grating made of Au, which optimized the synchrotron light flux in the range of 13–100 eV. The available experimental assembly included a differential pumping system that allowed the use of a Neon filter (0.65 mbar) to attenuate the X-ray portion of the synchrotron radiation spectrum (above 21.6 eV, or below 57.6 nm). This gives an effective energy range of 10–21.6 eV (124–57.6 nm) for the photons reaching the samples. The synchrotron emission reaches 400 nm, but as the emitted energy decays exponentially beyond this, this results in an effective cutoff of longer wavelengths above 124 nm (Fig. 1, see Section 3.1). Measurements of the photon flux were taken using a photodiode (IRD AXUV100).

Using the monochromatic beam and the Au grating, a graph of the current measured by the photodiode for each photon energy, I(E), was obtained. At the moment of each irradiation, the actual current that reached the sample can be calculated by this energy-dependent current, I(E), divided by the photodiode's quantum efficiency (QE$\left(\frac{electrons}{photons}\right)$), multiplied by the photon energy E(eV) to result in energy units. The result was then corrected with a scaling factor given by the measured polychromatic beam current at the time of the experiment ($I_{poly}$) divided by the integral of I(E) over the region of interest $I_{mono} = 9.2838 \times 10^{-9}$.

An aluminum mask was used to limit the irradiation to a 3 mm radius (R) circle of the sample, with the area A expressed in $m^2$.

From the photodiode manufacturer's calibration curve, the equation describing the behavior of its QE as a function of photon energy E(eV), from 10 to 21.6 eV, is given by: QE = 0.215 × E-1.05. It is assumed that the number of photons reaching the sample with energies below 10 eV is too small to be considered in the calculations.

Thus, the TGM flux R(E), in W/m$^2$ given by Eq. 1. Integrating from 10 to 21.6 eV as a function of energy, we obtain Eq. 2.

$$R(E)\left[Wm^{-2}\right] = \frac{I_{poly}}{I_{mono}}\frac{I(E)\,E}{QE(E)} \times \frac{1}{A}$$
$$= \frac{I_{poly}}{I_{mono}}\frac{I(E)\,E}{0.215\,E - 1.05}\frac{1}{\pi\left(R \times 10^{-3}\right)^2}\;;$$

(1)

$$E\left[Wm^{-2}\right] = \frac{I_{poly}}{9.2838 \times 10^{-9}}\,0.00204898\;;$$

(2)

Simplifying and multiplying by the irradiation time, t, we obtain the fluence in J/m² (Eq. 3).

$$F(I_{poly})\left[Jm^{-2}\right] = 0.2207\,I_{poly}(\mu A)t(s)\;;$$

(3)

The TGM flux in the VUV region (57.6–124 nm) was obtained from the measured current alongside the experiments, being 32.36 μA (mean value) and resulting in a mean value of 7.14 W/m². This was the mean VUV flux output used for our experiments.

### 2.4. Estimation of the irradiation parameters from the young Sun

2.4.1. Quiescent VUV emission. As stated earlier, we selected the 0.75 Gyr old solar analog κ¹ Cet as a representative for the young Sun at the time when life arose on early Earth. To obtain a spectrum ranging from the extreme ultraviolet (EUV) to the UV, thereby covering the VUV wavelength regime, different datasets were used. For the EUV range (∼10–90 nm), a synthetic spectrum based on coronal models was used (Sanz-Forcada and Ribas, 2015; Sanz-Forcada et al., 2011) because of the strong interstellar medium absorption from ∼40 nm up to ∼90 nm.

For the far-ultraviolet (FUV) and the UV, a merged spectrum that comprises the Far-Ultraviolet Spectroscopic Explorer (FUSE) and Hubble Space Telescope (HST) data covering the spectral range of 93–200 nm from Ribas et al (2010) was used. All these data were combined to construct a synthetic spectrum of κ¹ Cet ranging from the EUV to the UV. Finally, a wavelength-integrated quiescent flux of κ¹ Cet in the spectral region 57.6–124 nm, corresponding to the VUV range of the synchrotron, was determined from the reconstructed spectrum, and scaled to the orbits of different potentially habitable objects in the solar system (Venus, Earth, and Mars; cf. Section 3.1).

2.4.2. VUV emission during flares and superflares. To estimate how flaring would enhance the flux of κ¹ Cet relative to its quiescent value, we analyzed observations of the Extreme Ultraviolet Explorer (EUVE) satellite (this is the only satellite that was operating at those wavelengths), which captured a number of flares on late-type main sequence stars, mainly in its deep survey (DS) channel, which collects photons in the range 8–18 nm. A visual inspection of those light curves from the literature (e.g., Audard et al., 2000; Sanz-Forcada and Micela, 2002; Sanz-Forcada et al., 2003) reveals that peak enhancements up to one order of magnitude was observed, predominantly for dM stars. K¹ Cet has also been observed by EUVE during a flare displaying a peak enhancement factor of ∼2 (see

Fig. 2 in Section 3.1). The DS band covers shorter wavelengths than the 57.6–124 nm range used in our study. Therefore, these enhancement factors represent rough estimates only.

There is only a small number of flare observations of young main sequence stars in the EUV, a fact related to the difficulties of observing at EUV wavelengths because of interstellar absorption and the fact that there was only one satellite (EUVE) operating at those wavelengths. Therefore, to estimate possible flare flux enhancement factors in the VUV range considered here, one needs to use adjacent wavelength regimes at which more observatories operated and from which a higher number of flares than in the EUV was detected. Karmakar et al. (2017) presented superflares from CC Eri, a nearby binary system (K7.5Ve + M3.5Ve) observed by Swift.

The authors found a peak luminosity enhancement of ∼3500 during a flare, a factor that we adopted as the maximum possible enhancement during superflares. Güdel et al. (2004) found peak flux enhancements of >100 for flares on the mid-M dwarf Proxima Cen observed in X-rays using XMM-Newton. Horan and Kreplin (1981) presented simultaneous observations of solar flares at EUV and soft X-ray wavelengths. These authors found a difference up to a factor of three in flare emissions in both wavelength regimes, suggesting a similar flare emission level at both wavelength regimes of the Sun.

However, the stars mentioned earlier are not strictly solar-like, so we use these results with caution. Therefore, we consider enhancement factors obtained from X-ray investigations during stellar flaring to be upper limits. We want to stress that the estimation of the flare enhancement factor in the VUV range can be taken just as an order of magnitude estimation.

2.4.3. Occurence rates of superflares. There have been a few observations of superflares from κ¹ Cet. One is related to an HeI D3 line enhancement seen in optical spectral observations (Robinson and Bopp, 1987) with an estimated flare energy of ∼10³⁴ erg (10²⁷ J) (Schaefer et al., 2000). Two others are EUV flares captured by EUVE (see also Fig. 2, in Section 3.1) and presented in the work of Audard et al. (2000) with flare energies of ∼10³³ erg (10²⁶ J) and ∼2 × 10³³ erg (2 × 10²⁶ J).

To estimate the occurrence rate of superflares from κ¹ Cet, we use the fact that flare frequencies have the form of a power-law distribution as follows:

$$dN/dE \propto E^{-\alpha};$$

(4)

Here, E is the flare energy and α the power-law index. Hence, the cumulative frequency of flares with energies >E can be written as $N(>E) \propto E^{-\alpha+1}$.

By knowing α and the typical flare frequency at a certain energy, one can extrapolate the occurrence rate of larger flares under the assumption that the power law remains valid up to these energies. The maximum flare energy that can be generated depends mainly on the size and magnetic field strength of a starspot group (Shibata et al., 2013). By using Eq. 1 of Shibata et al., together with the maximum magnetic field strength observed for a sunspot (∼6000 G) (Okamoto and Sakurai, 2018) and the maximum spot group area found in solar-like stars (∼0.1 stellar hemispheres, see Fig. 2 of Shibata $et\ al.$, 2013), we estimated the maximum attainable flare energy to be of the order of $10^{30}$ J for a solar-like star.

This estimate is comparable to the maximum observed bolometric flare energies for young solar-like stars from $TESS$ data (Tu $et\ al.$, 2020). At this point, we stress that it is unknown whether $\kappa^1$ Cet or the young Sun ever had superflares up to the maximum flare energy estimated earlier, and therefore the extrapolation to such high energies for any specific star is just a working hypothesis. Based on the close correlation between the stellar X-ray luminosity and the cumulative frequency of flares with XUV (X-ray and EUV, 0.01–10 keV) energies above $10^{32}$ erg ($10^{25}$ J) (Audard $et\ al.$, 2000), we can estimate the frequency of XUV superflares ($E > 10^{26}$ J) using the relation:

$$N(> E)/N(> 10^{25}) = (E/10^{25})^{-\alpha+1};\qquad (5)$$

This Eq. 5 was derived from Eq. 4. The resulting extrapolated superflare occurrence rates of $\kappa^1$ Cet are shown in Section 3.1.

### 2.4.4. Connection with VUV experiments.

To define the exposure time for the VUV irradiations, we calculated the fluence [$i.e.$, in photobiology and defined according to Braslavsky (2007) at a given point in space, the radiant energy incident on a small sphere from all directions divided by the cross-sectional area of that sphere, SI unit J/m$^2$]. In the case of our study, to calculate the fluence we considered the energies emitted by superflares. The flare energy E can be approximated as:

$$E = (F - F_Q)\ \Delta t\ 2\pi a^2;\qquad (6)$$

Here, F is the flux at the planetary orbit during flaring conditions, $F_Q$ is the flux during quiescent conditions, and $\Delta t$ is the duration of the flare. Here, we neglect the light curve shape of a typical flare and approximate it with a constant value of F, which is an adequate approximation close to the flare peak. The term $2\pi a^2$ scales the flare radiation emitted by the star into the visible hemisphere ($2\pi$) to the orbital distance "a" of the planet. Setting the flare flux F equal to the synchrotron flux used in the experiments, we can simplify Eq. 6 to $E \approx F\ \Delta t\ 2\pi a^2$, as the quiescent stellar fluxes $F_Q$ are very small compared with F (cf. Section 3.1).

For the orbital distances considered (Venus, Earth, and Mars), we estimate that flare durations of $\Delta t > 4$ min correspond to VUV flare energies that are $\geq 10^{26}$ J, that is, superflares. This $\Delta t$ is comparable to typical peak durations of X-ray flares, which are in the order of a few minutes to hours (Pye $et\ al.$, 2015).

In addition, to establish a relation between flare energies, the fluences they would generate at different planetary orbits, and the fluences used in the experiments, we define the fluence as $f = F\ \Delta t$, and therefore the flare energy is related to the fluence via:

$$E = f \times 2\pi a^2;\qquad (7)$$

### 2.5. Flare simulation and VUV irradiation experiments

The microorganisms (prepared according to Section 2.2) were introduced in a sample holder into a vacuum chamber and exposed to a decreasing vacuum pressure for 3 h, which was the time required to reach a high vacuum (in our experiment this value was $1 \times 10^{-6}$ mbar $= 10^{-4}$ Pa). This low pressure was maintained throughout the entire irradiation experiment (around 15 h).

Inside the same chamber, the microorganisms (prepared according to Section 2.2) were bombarded with VUV radiation with different exposure times (between 4 and 5400 s) to represent the duration of a flare or a superflare ($e.g.$, Pye $et\ al.$, 2015). We additionally included shorter times to probe a range of different fluences of 30, 100, 300, 1000, 2000, 2500, 5000, 10,000, 20,000, and 40,000 J/m$^2$ (Table 1). The experiments were performed at room temperature. The irradiations were done in triplicate, and the control groups were obtained for each kind of condition in parallel (non-irradiated samples).

We additionally checked to discern whether UV wavelengths longer than 124 nm emitted by the TGM synchrotron would have any incidence in the survival of the microorganisms. For that, we made a comparison with the survival curves to UV obtained for air-dried $D.\ radiodurans$ R1 cells by Bauermeister $et\ al.$ (2011) and with those obtained for halite-embedded cells of several haloarchaeal strains (Fendrihan $et\ al.$, 2009). According to our estimated fluences and the survival curves reported in these studies, the contribution of UV (200–400 nm) would not have any significant effects on the survival of the microorganisms in our experiments. Therefore, we did not consider the influence of wavelengths above 124 nm in our study.

### 2.6. Recovery and survival analysis

The microorganisms were recovered after the irradiation experiment by adding 10 μL of culture medium to the spots of microorganisms on the glass slides and then mixed with a micropipette for 10 s to achieve maximum recovery of the cells. Aliquots were seeded on agar plates in spots by triplicate (without dilution and 1/100 dilution) and incubated. Then, we determined the number of viable cells (viable count) by their ability to form colonies; this was counted as colony-forming units per μL (CFU/μL). The limit of detection (LOD) was determined to be 10 CFU/μL, which corresponds to a count of 10 colonies in an undiluted sample.

The survival fractions were calculated generically with the fraction $N/N_0$, where $N$ is the viable count after treatment and $N_0$ is the viable count before treatment. In the case of our study, we had multiple consecutive steps of exposure to different conditions, different $N$ and $N_0$, described with different letters according to the treatment. $N_D$ corresponds to microorganisms that were deposited in glass slides and air-dried, but not exposed to vacuum nor to radiation; $N_V$ corresponds to microorganisms deposited in glass slides (following the same procedure as for $N_D$) that were exposed to vacuum alone; and $N_R$ corresponds to microorganisms deposited in glass slides (following the same procedure than for $N_D$) but that was exposed to vacuum and VUV radiation simultaneously.

The survival fraction to vacuum exposure was calculated simply as $N_V/N_D$. The survival fraction to radiation was determined by comparing $N_R$ (CFU/μL) after exposure with different fluences with $N_V$. The survival curves to VUV were obtained by plotting the $N_R$ as a function of the fluence of the $y$-intercept yielding the $N_V$ (control group cells that survived in the vacuum and had not been exposed to VUV). These curves were then analyzed using the single-hit target theory model (Lea, 1946), as follows:

$$\log N_R = \log(N_V e^{-kHo}). \tag{8}$$

In Eq. 8, Ho is the applied fluence, and "k" in the exponential is an inactivation coefficient equal to the reciprocal of the fluence required to reduce the survival fraction $N_R/N_V$ to 1/e (37%). The parameters of the equation were adjusted to the experimental data by a non-linear regression. This model was not suitable to describe the response of naked cells of *D. radiodurans* in our experiments, because the slope of the survival curve decreased at high fluences. In this case, an alternative model was used, which assumes the existence of a minor subpopulation of microorganisms able to tolerate high doses of radiation (Hiatt, 1964). The equation corresponding to this model is:

$$\log N_R = \log(N_V (f e^{-k1Ho} + (1-f) e^{-k2Ho})); \tag{9}$$

where "k1" in the first exponential is the inactivation coefficient for the susceptible subpopulation of microorganisms, which represents a majority fraction f of the total population of microorganisms, and "k2" is the inactivation coefficient for the tolerant subpopulation of microorganisms.

The Ho$_{90}$ quantity was also calculated, corresponding to the fluence required to reduce the survival fraction by 90% when the microorganisms are exposed to VUV + vacuum.

### 2.7. Microscopy analysis

Additional independent experiments were conducted to study the crystal morphology and distribution of the deposited cells or the embedding of cells inside salt crystals, and the identification of eventual cell membrane damage before the exposure to vacuum and VUV radiation. Aliquots of 50 μL of all suspensions prepared for the conditions described in Section 2.2 were stained with LIVE/DEAD® *Bac* Light™ Bacterial Viability Kit (Life Science), which uses a mixture of a green-fluorescent nucleic acid stain (SYTO®9), and a red-fluorescent nucleic acid stain, propidium iodide (PI). Simultaneous application of both dyes in the samples results in differential staining as PI penetrates only when the outer cell membrane is damaged (PI signal, red fluorescence), whereas SYTO 9 generally labels all bacteria in a population (SYTO 9 signal, green fluorescence) (Stocks, 2004).

The samples in the present study were stained according to manufacturer instructions with modifications (Leuko *et al.*, 2004). After that, the same evaporation procedure described in Section 2.2 was performed. Naked cells and salt or halite-embedded cells of microorganisms were observed with an Olympus FluoView™ FV300 confocal laser microscope using a sequential sweep with the following settings. Channel 1 (SYTO 9): excitation 488 nm, emission filter 510–530 nm. Laser intensity 5%, PMT 650 V, gain 1, offset 0. Channel 2 (PI): excitation 543 nm, emission filter 585–620 nm. Laser intensity 50%, PMT 650 V, gain 1, offset 0. Air immersed 4×(0.1 NA), 20×(0.5 NA), or oil-immersed 60×(1.40 NA), 100×(1.40 NA) objectives were used.

The images obtained by confocal microscopy were analyzed using the image processing package FIJI/Image J (Schindelin *et al.*, 2012). The morphology of salt crystals and halite crystals was also studied by contrast phase using optical microscopy (Olympus BX51P microscope).

## 3. Results

### 3.1. VUV radiation from the young Sun

The comparisons between the reconstructed quiescent spectrum of κ¹ Cet (young Sun; see Section 2.4.1) with the quiescent spectrum of the present Sun (both scaled to an orbital distance of 1 AU) and the TGM synchrotron beamline are shown in Fig. 1. This plot shows that the synchrotron flux is between two and more than three orders of magnitude higher than the quiescent spectral flux of the young Sun, particularly for the wavelength region from 57.6 to 90 nm.

In addition, the magnetic activity of the young Sun is expected to have been higher than that of the present Sun for the radiation emitted below 170 nm, given the evidences provided by the solar analog κ¹ Cet. Therefore, the quiescent VUV flux of the young Sun would have been between one to two orders of magnitude higher than that of the present Sun.

From the reconstructed spectrum shown in Fig. 1, the VUV flux was obtained by integrating over the corresponding wavelength range. The resulting quiescent VUV flux of κ¹ Cet was then scaled to the orbits of potentially habitable planetary bodies in the early Solar System 3.8 Gyr ago, which yielded values of 0.088, 0.046, and 0.02 W/m², respectively, for the orbits of Venus, Earth, and Mars. The estimated mean VUV flux value emitted by the TGM synchrotron during our experiments was 7.14 W/m² (see Section 2.3).

When comparing this value with the quiescent VUV fluxes at different planetary orbits, it can clearly be seen that the VUV flux emitted by the TGM synchrotron is much higher than the quiescent emission of the young Sun in this wavelength region. Thus, the experiments are *not* representative of the quiescent conditions in the young solar system but of conditions with strongly enhanced flux levels, such as during stellar flares or superflares.

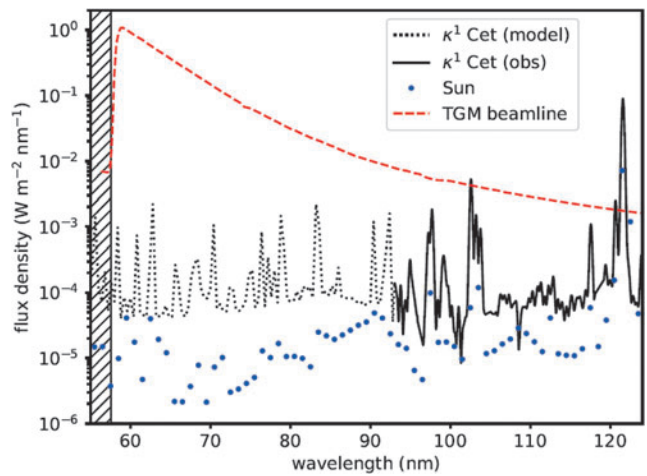

**FIG. 1.** Spectral flux in the VUV range (57.6–124 nm) scaled to an orbital distance of 1 AU for the present Sun and κ¹ Cet (young Sun), compared to the TGM synchrotron beamline. The hatched area on the left indicates X-ray and VUV wavelengths below 57.6 nm that were attenuated during the experiments. TGM, toroidal grating monochromator; VUV, vacuum-ultraviolet.

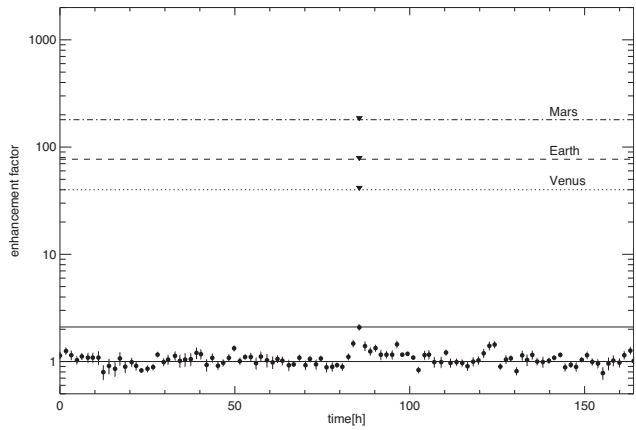

**FIG. 2.** The strongest flare detected by EUVE (8–18 nm) of the solar analog κ[1] Cet (black dots). The peak of the flare (thick black solid line) was enhanced with respect to the quiescent level (thin black solid line) by a factor of ∼2.2. For comparison, to obtain a flux similar to the TGM flux used in the experiments, κ[1] Cet's quiescent emission would have to be raised by factors 80–360 depending on the orbits of Venus, Earth, and Mars, which could only be achieved by very energetic superflares. The enhancement factors are shown as horizontal lines of different styles and inverted triangles, indicating the required enhanced flare peak at the corresponding planetary orbit. EUVE, Extreme Ultraviolet Explorer.

Considering the different planetary orbits, different flux enhancement factors with respect to the quiescent level are required to reach the flux value of the experiments (i.e.,

TABLE 1. FLARE ENERGIES GENERATING HYPOTHETICAL VACUUM-ULTRAVIOLET FLUENCES AT THE ORBITS OF EARLY PLANETARY BODIES OF THE SOLAR SYSTEM (VENUS, EARTH AND MARS) THAT ARE COMPARABLE TO THOSE FROM THE EXPERIMENTS AT THE TOROIDAL GRATING MONOCHROMATOR

| *VUV fluences ($J/m^2$) used in the experiments* | *Calculated flare energies (J) corresponding to different planetary orbits* | | |
|---|---|---|---|
| *TGM* | *Venus* | *Earth* | *Mars* |
| 30 | $2.1 \times 10^{24}$ | $4.2 \times 10^{24}$ | $9.5 \times 10^{24}$ |
| 100 | $6.9 \times 10^{24}$ | $1.4 \times 10^{25}$ | $3.2 \times 10^{25}$ |
| 300 | $2.1 \times 10^{25}$ | $4.2 \times 10^{25}$ | $9.5 \times 10^{25}$ |
| 1000 | $6.9 \times 10^{25}$ | $\mathbf{1.4 \times 10^{26}}$ | $\mathbf{3.2 \times 10^{26}}$ |
| 2000 | $\mathbf{1.4 \times 10^{26}}$ | $\mathbf{2.8 \times 10^{26}}$ | $\mathbf{6.4 \times 10^{26}}$ |
| 2500 | $\mathbf{1.7 \times 10^{26}}$ | $\mathbf{3.5 \times 10^{26}}$ | $\mathbf{8.0 \times 10^{26}}$ |
| 5000 | $\mathbf{3.4 \times 10^{26}}$ | $\mathbf{7.0 \times 10^{26}}$ | $\mathbf{1.6 \times 10^{27}}$ |
| 10,000 | $\mathbf{6.9 \times 10^{26}}$ | $\mathbf{1.4 \times 10^{27}}$ | $\mathbf{3.2 \times 10^{27}}$ |
| 20,000 | $\mathbf{1.4 \times 10^{27}}$ | $\mathbf{2.8 \times 10^{27}}$ | $\mathbf{6.4 \times 10^{27}}$ |
| 40,000 | $\mathbf{2.8 \times 10^{27}}$ | $\mathbf{5.6 \times 10^{27}}$ | $\mathbf{1.3 \times 10^{28}}$ |

Energies $\geq 10^{26}$ J would correspond to superflares and are marked in bold.
TGM = toroidal grating monochromator; VUV = vacuum-ultraviolet.

the more distant the planet, the higher the flux enhancement has to be to reach a value of 7.14 W/m²). The corresponding flare enhancement factors are about 80, 155, and 360, for the orbits of Venus, Earth, and Mars, respectively.

This range of flux enhancement factors between 80 and 360 is comparable to stellar flare peak enhancement factors observed in EUV and X-rays found in the literature for very energetic superflares (cf. Section 2.4.2; see a graphic example in Fig. 2). Therefore, a stellar superflare could provide such enhancement factors of the VUV flux levels, and the radiation emitted by the TGM synchrotron can be used to simulate the flux emitted by a stellar superflare event.

Considering a superflare VUV peak flux value of 7.14 W/m², we selected different irradiation times for the experiments in the present study that would mimic the duration of the superflares, and from that, we obtained different fluences as shown in Table 1. Taking radiation times for flare peaks lasting from minutes to hours into account (e.g., Pye et al., 2015), we obtained values between 1000 and 40,000 J/m². Fluences between 30 and 300 J/m² would correspond to flare durations of a few seconds; we included such low fluences in our experiments to get a broader picture of the biological response of the microorganisms to VUV.

In agreement with these times, it is also possible to establish which stellar flare energies would be required to generate VUV fluences comparable to those used in our experiments at the respective planetary orbits (see Section 2.4.4, Eq. 7). The results are shown in Table 1.

After establishing that the experimental parameters (fluxes, fluences) are comparable to stellar superflares, the question remains as to how common such superflares were in the early Solar System. From the two sets of EUVE observations of κ[1] Cet that each includes a superflare, Audard et al. (2000) obtained superflare rates of about 2 and 5 per day, respectively. However, the total observing time was of the order of 2 weeks and, therefore, did not include the rarer, more energetic superflares, in case they exist. On the other hand, an estimation of the total superflare frequency $N(E > 10^{26}$ J) of κ[1] Cet using Eq 5 yields a frequency of ∼7 per day, that is, slightly higher than the observed but still comparable.

For an estimation of the superflare frequency with Eq. 5, all parameters were taken from the work of Audard et al. (2000). For $N(>10^{25})$ (number of flares per day with XUV energies $10^{25}$ J and higher), we use the relation of this quantity with X-ray luminosity $L_X$ (∼$10^{22}$ W for κ[1]Cet) as already mentioned in Section 2.4.3. For the flare power-law index α, we use the weighted average of the values determined from the two EUVE data sets of κ[1] Cet, which gives $2.26 \pm 0.44$. The error on $N(>E)$ is dominated by the uncertainty of α, so we ignore the other contributions.

Figure 3 shows the extrapolated relation between the XUV superflare energy E and the cumulative number of superflares above E per year (solid line) and associated uncertainty range due to α (dashed lines). We also show the time in years to undergo one superflare with an XUV energy >E. The uncertainty of this extrapolation spans up to several orders of magnitude and increases with superflare energy. As mentioned earlier, just two XUV superflares were observed by EUVE on κ¹ Cet. These observations yield a cumulative superflare rate of ~0.2 per day with energies >$10^{26}$ J (see Fig. 2 in Audard et al, 2000), which corresponds to about 70 superflares per year, which is consistent with the extrapolated range using Eq. 5 (50–400 superflares per year with E > $10^{26}$ J; Fig. 3).

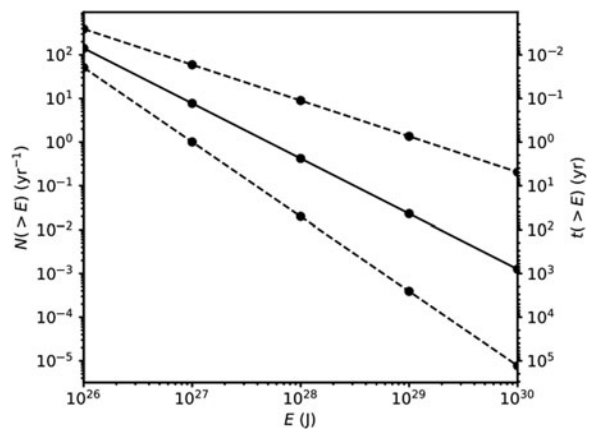

**FIG. 3.** Extrapolation of the cumulative number of super-flares with XUV energies >E per year (left *y* axis) and corresponding time to have one superflare with an XUV energy >E (right *y* axis) for κ ¹Cet. The solid line shows the average relation and the dashed lines its error range due to the uncertainty of the flare power-law index α.

### 3.2. Effects of air-drying and halite entrapment in D. radiodurans

Suspensions of *D. radiodurans* in distilled water and in 4 *M* NaCl were deposited on the top of glass slides and air-dried. Naked cells and halite crystal embedded cells were obtained, respectively (Fig. 4). The number of viable cells for each condition ($N_D$) is shown in Table 2. Given that the cell membrane has been considered classically a main target for VUV damage and the distribution of the cells is relevant for VUV shielding, we evaluated the cell membrane status and distribution of microorganisms during the sample preparation and before exposure to vacuum and VUV radiation. To this end, an additional experiment was performed where the cells were resuspended in distilled water and 4 *M* NaCl, and they were co-stained with PI and SYTO 9 (LIVE/DEAD *Bac*Light Bacterial Viability Kit) and analyzed under confocal microscopy.

The air-dried microorganisms in distilled water exhibited naked cells that were grouped, though distributed in mono-layers (Fig. 4). The composite image shows that most of the naked cells have signals for both PI and SYTO 9, which could indicate that the cell membranes of *D. radiodurans* are slightly or partially damaged (PI partially entered the cells, whereas SYTO 9 is always present) (Fig. 4A, B). Cells with partially damaged membranes, in general, may contain different amounts of PI entering the cells (different proportions of PI and SYTO 9 signal are ascribed to orange and yellow fluorescence, see, *e.g.*, Boulos *et al.*, 1999).

A very low fraction of cells shows a complete PI signal, which can be attributed to total cell membrane damage (see Fig. 4C, examples indicated by arrows). This is confirmed by the images that show the fluorescence channels sepa-rately for SYTO 9 (Fig. 4C) or PI (Fig. 4D), where it is possible to see that almost all the cells contain both SYTO 9 and PI signals.

The air-dried microorganisms in 4 *M* NaCl showed halite crystal-embedded cells. The microorganisms were not clumped and distributed inside fluid inclusions within the crystal structure. There were no cells identified in the crystal matrix outside the fluid inclusions (Fig. 5). The composite image (Fig. 5A) shows mostly a SYTO 9 signal and very few cells with a combination of SYTO 9 and PI signals (meaning that PI did not completely displace SYTO 9).

This could indicate that the exposure to 4 *M* NaCl (entrapment inside halite) would have almost no effect on the cell membrane integrity for *D.* radiodurans, as there were not enough amounts of PI inside the cell to displace SYTO 9. This is confirmed by images that show the fluorescence channels separately only for SYTO 9 (Fig. 5B) or PI (Fig. 5C), where it is possible to see that most of the cells have SYTO 9 but no PI signal.

### 3.3. Effects of vacuum in D. radiodurans

Naked cells and halite crystal-embedded cells were exposed to vacuum conditions inside a vacuum chamber until reaching a constant pressure value of $10^{-4}$ Pa (equivalent to pressure values in the low Earth orbit) for around 15 h. The initial number of viable cells for each condition exposed to vacuum ($N_V$) and the survival fraction to vacuum exposure ($N_V/N_D$) are shown in Table 2.

Our results show a decrease of four orders of magnitude in the survival for naked cells when exposed to vacuum and a decrease in the survival between two and three orders of magnitude in the case of halite crystal-embedded cells.

The comparison between the survival fractions to vacuum shows that halite crystal-embedded cells have a survival fraction at least one order of magnitude higher than naked cells, revealing a protective effect for cells inside fluid inclusions of halite when exposed to the effects of vacuum.

### 3.4. Effects of vacuum plus VUV radiation in D. radiodurans

Naked cells and halite crystal-embedded cells were exposed to different fluences of VUV radiation plus vacuum to simulate the exposure to flaring conditions in the VUV range emitted by the young Sun.

The viable number of cells for each fluence and condition was obtained ($N_R$) and plotted as a function of the VUV fluence (Fig. 6).

The survival curve for naked cells shows a bimodal behavior, with a rapid decrease in the survival for fluences up to 300 J/m² (which would correspond to the case of a flare in the orbit of Venus, Earth, or Mars, see Table 1). Above this value, there is a tail in the survival curve up to 2000 J/m² that would indicate there is a fraction of microorganisms (between 10 and 100 CFU/sample) capable of surviving higher fluences, which would correspond to flares or superflares depending on the planetary orbit (Table 1).

In contrast, the results for halite crystal-embedded cells show a monotonic exponential decrease in survival up to a fluence of 5000 J/m², which would correspond to a superflare for all considered planetary orbits.

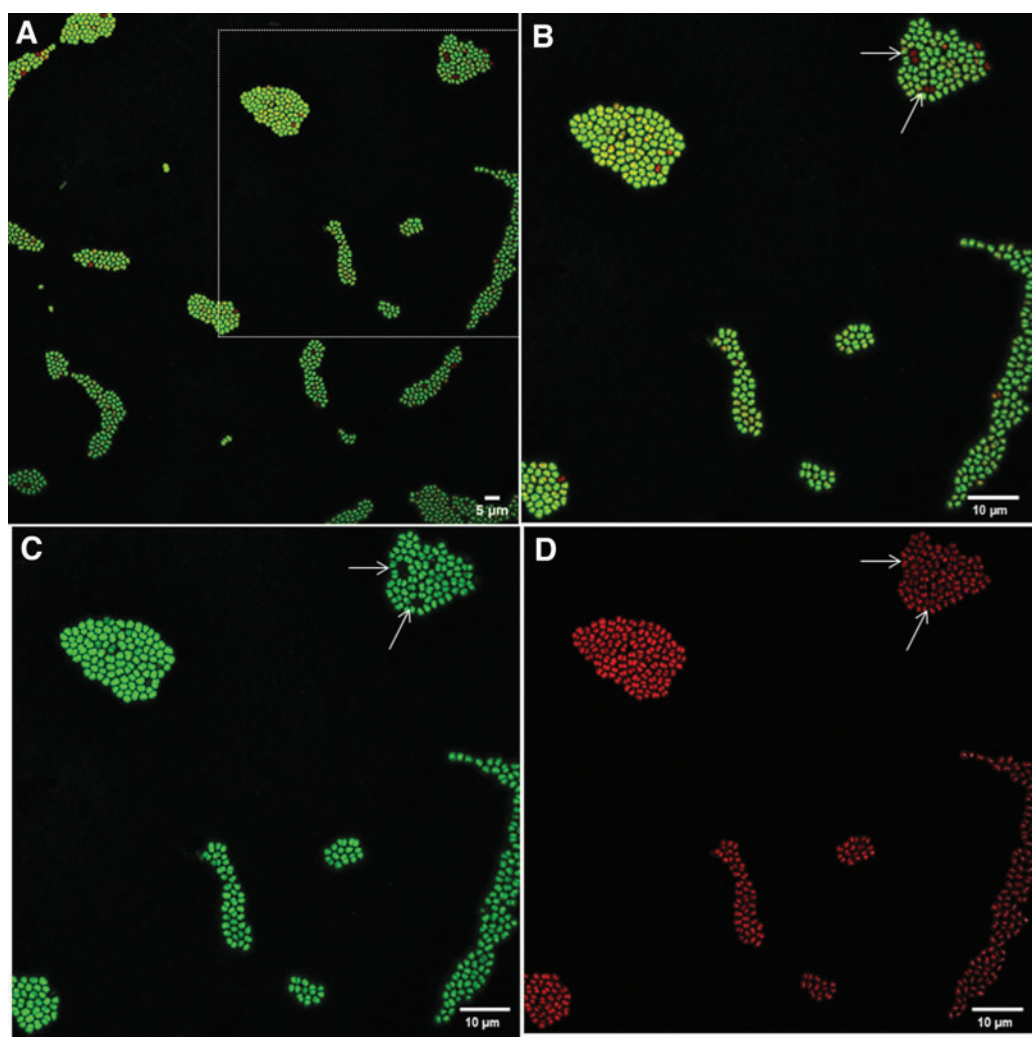

**FIG. 4.** Confocal microscopy images of *Deinococcus radiodurans* desiccated in distilled water co-stained with SYTO 9 and PI showing characteristics of the distribution of microorganisms and potential damage to the cell membrane. **(A)** Color composite (SYTO 9 and PI), 1000×magnification. **(B)** Amplification of a 100×100 µm section of **(A)** (indicated as a dotted square in **(A)**, as **(C, D)** SYTO 9, and PI signal, respectively. The arrows indicate a sample of cells with PI signal but without SYTO 9 signal. PI, propidium iodide.

Although a few halite crystal-embedded cells survived fluences up to 10,000 J/m², the corresponding data are not shown in Fig. 6 nor are they included in the regression analysis (see Section 4) as the viable counts at high doses were below the LOD. The analysis of the initial slopes of the curves (*k*-value, Table 3) shows a high decrease in the *k*-value for halite crystal-embedded cells, in comparison with the *k*-value for naked cells, which can be attributed to the protective effects of the halite to VUV. This can be more easily visualized through the comparison of the $Ho_{90}$

TABLE 2. NUMBER OF VIABLE CELLS (EXPRESSED AS CFU/µL) FOR AIR-DRIED MICROORGANISMS ($N_D$) AND AIR-DRIED PLUS VACUUM TREATED MICROORGANISMS ($N_V$)

| Microorganism | Condition | $N_D \pm SD$ (CFU/µL) | $N_V \pm SD$ (CFU/µL) | $N_V/N_D \pm SD$ (%) |
|---|---|---|---|---|
| *D. radiodurans* | Naked cells | $1.28 \times 10^6 \pm 4.80 \times 10^5$ | $2.25 \times 10^2 \pm 3.27 \times 10^1$ | $1.75 \times 10^{-4} \pm 2.55 \times 10^{-5}$ (0.017%) |
| *D. radiodurans* | Halite crystal-embedded cells | $1.93 \times 10^5 \pm 2.52 \times 10^4$ | $1.61 \times 10^3 \pm 1.24 \times 10^3$ | $8.33 \times 10^{-3} \pm 6.44 \times 10^{-3}$ (0.83%) |
| *H. volcanii* | Salt crystal-embedded cells | $4.03 \times 10^6 \pm 8.13 \times 10^5$ | $1.51 \times 10^4 \pm 2.30 \times 10^3$ | $3.74 \times 10^{-3} \pm 5.71 \times 10^{-4}$ (0.37%) |
| *H. volcanii* | Halite crystal-embedded cells | $2.27 \times 10^5 \pm 1.89 \times 10^4$ | $2.83 \times 10^3 \pm 1.48 \times 10^3$ | $1.25 \times 10^{-2} \pm 6.53 \times 10^{-3}$ (1.25%) |

Survival fraction to vacuum is shown as $N_V/N_D$. The corresponding percentage of survival is indicated between parentheses.
CFU = colony-forming units; SD = standard deviation.

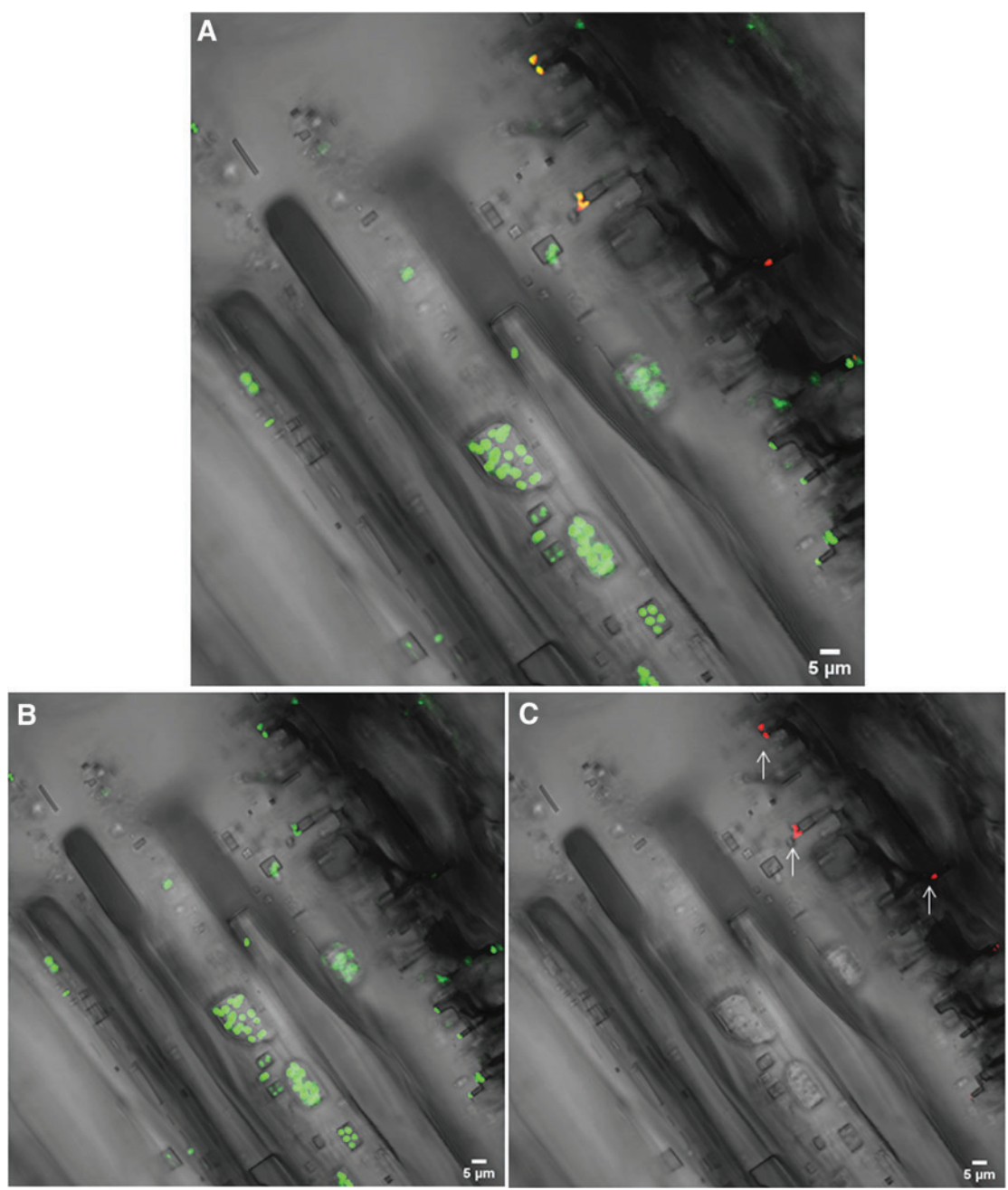

**FIG. 5.**    Confocal fluorescence microscopy images (1000×) of halite crystal-embedded cells of *Deinococcus radiodurans* co-stained with SYTO 9 and PI, showing characteristics of the crystal, the distribution of microorganisms inside fluid inclusions, and potential damage to the membrane. **(A)** Color composite (SYTO 9 and PI). **(B, C)** SYTO 9 and PI signal, respectively. The arrows indicate cells with PI signal without SYTO 9 signal.

(Table 3) value, which shows that it would be necessary to apply 11 times more fluence to kill most of the initial population of microorganisms if they are embedded in halite crystals than if they are not (Table 3).

### 3.5.  Effects of salt crystal and halite crystal entrapment in H. volcanii

Suspensions of *H. volcanii* in saline solution and in 4 *M* NaCl were deposited on top of glass slides and air-dried. Salt crystal-embedded cells and halite crystal-embedded cells were obtained, respectively (Fig. 7). The number of viable cells for each condition ($N_D$) is shown in Table 2. Following the steps outlined in Section 3.2, we evaluated the cell membrane status and distribution of microorganisms during sample preparation and before the exposure to vacuum and VUV radiation, where cells were stained (LIVE/DEAD *Bac*Light Bacterial Viability Kit) and analyzed under confocal microscopy.

The results for air-dried cells in saline solution show that the evaporation leads to the formation of salt crystals. Inside the crystals, the microorganisms were not clumped and distributed inside fluid inclusions within the crystal structure. There were no cells identified in the crystal matrix outside the fluid inclusions (Fig. 7). The composite images show that most of the cells exhibit SYTO 9 signal, which indicates that cell membranes were not damaged during the process of entrapment (Fig. 7A). This is confirmed by the images in separate channels, which show that there is mainly SYTO 9 signal (Fig. 7B) and that a few cells present PI signal, which would have cell membrane damage (Fig. 7C).

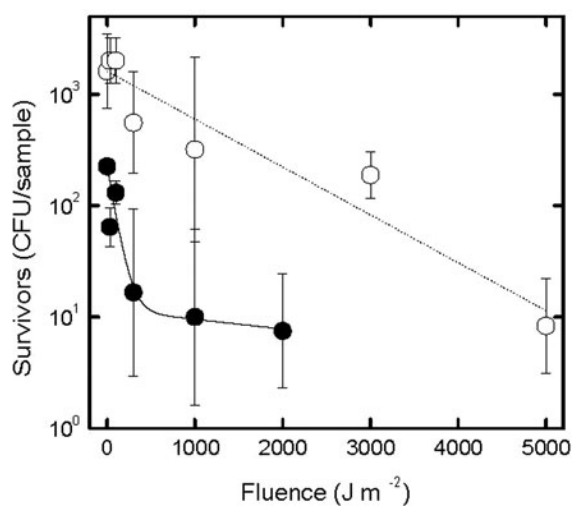

**FIG. 6.** *Deinococcus radiodurans* naked cells (black circles) and halite crystal-embedded cells (white circles) exposed to different fluences of VUV plus vacuum ($N_R$). Vertical bars indicate standard errors. Dotted lines represent the adjustment to non-linear regression.

The results for air-dried cells in 4 *M* NaCl show that the evaporation led to the formation of halite crystals and that the cells were distributed inside fluid inclusions without clumping inside the crystal. The composite images (Fig. 8A) show that most of the cells have SYTO 9 signal, which indicates that the process of entrapment inside the halite did not affect the membrane integrity. This is confirmed by the images shown in separate channels, which show that only SYTO 9 signal is present (Fig. 8B) and that the PI signal is only seen in a few cells (Fig. 8C).

Even though both saline solution and 4 *M* NaCl lead to the formation of salt crystals, they show distinctive characteristics, and their structure seems to be different (Figs. 7 and 8). This was confirmed by additional optical microscopy observations (Fig. 9). In general, halite crystals are of bigger size than salt crystals by 3500–5500 µm (Fig. 9A) and 400–

800 µm (Fig. 9B), respectively. Moreover, halite crystals present a well-developed typical cubic crystal structure with a transparent appearance and with small (<20 µm) and big (~100 µm) fluid inclusions (Fig. 9A1, A2).

Salt crystals exhibit a hopper-like morphology, commonly attributed to crystallization under supersaturated conditions (*e.g.*, a rapid crystallization, Roedder, 1984), and present a skeletal structure with a ''dirty'' appearance (Fig. 9B). The opacity observed in the crystals is due to light scattering of numerous minuscule fluid inclusions (*e.g.*, 8 µm), along with others of bigger size (up to around 70 µm) (Fig. 9B1, B2) that comprise an extensive organized network of tubules.

Moreover, the cells adopt a different morphology inside the fluid inclusions, depending on whether they are entrapped within crystals of saline solution or in halite crystals: in the first case, it is possible to see rod-shaped cells (*bacilli*) (Fig. 7A), whereas in the second case, the cells appear as spherical-shaped (*cocci*) (Fig. 8A).

### 3.6. Effects of vacuum in H. volcanii

As in previous experiments, salt crystal-embedded cells and halite crystal-embedded cells were exposed to vacuum conditions inside a vacuum chamber until reaching a constant pressure value of $10^{-4}$ Pa (equivalent to pressure values in the low Earth orbit). The number of viable cells for each condition exposed to vacuum ($N_V$) and the survival fraction to vacuum ($N_V/N_D$) are shown in Table 2.

Our results show a decrease of three orders of magnitude in the survival for salt crystal-embedded cells and a decrease of two orders of magnitude for halite crystal-embedded cells that were exposed to vacuum.

The comparison between the survival fractions and vacuum shows that halite crystal-embedded cells have a survival fraction that is one order of magnitude higher than that of salt crystal-embedded cells. This reveals again a protective effect for cells inside fluid inclusions of halite when exposed to this condition.

Although in both cases the microorganisms were trapped inside fluid inclusions of crystal structures, halite crystals seem to provide better protection to vacuum.

### 3.7. Effects of vacuum plus VUV in H. volcanii

Salt mixture crystal-embedded cells and halite crystal-embedded cells were exposed to different fluences of VUV radiation plus vacuum, which corresponds to fluences from stellar flares or superflares. The viable number of cells for each fluence and condition was obtained ($N_R$) and plotted as a function of the VUV fluence (Fig. 10).

TABLE 3. PARAMETERS OBTAINED FROM THE VACUUM-ULTRAVIOLET SURVIVAL CURVES IN FIGS. 6 AND 10, WHERE $\kappa$ IS THE INITIAL SLOPE OF THE CURVE OF THE NUMBERS OF SURVIVORS AND $Ho_{90}$ CORRESPONDS TO THE FLUENCE WHERE 90% OF VIABLE CELLS FROM THE INITIAL POPULATION IS OBTAINED

| Microorganism | Condition | k Value ± SD[a] | $Ho_{90}$ ($J/m^2$) |
|---|---|---|---|
| D. radiodurans | Naked cells | $1.10 \times 10^{-2} \pm 5.15 \times 10^{-3}$ | 209.32 |
| D. radiodurans | Halite crystal-embedded cells | $9.90 \times 10^{-4} \pm 1.00 \times 10^{-3}$ | 2325.81 |
| H. volcanii | Salt crystal-embedded cells | $3.14 \times 10^{-3} \pm 2.20 \times 10^{-4}$ | 733.30 |
| H. volcanii | Halite crystal-embedded cells | $2.30 \times 10^{-3} \pm 6.00 \times 10^{-5}$ | 1001.11 |

[a]SD of *k* value obtained in the regression.

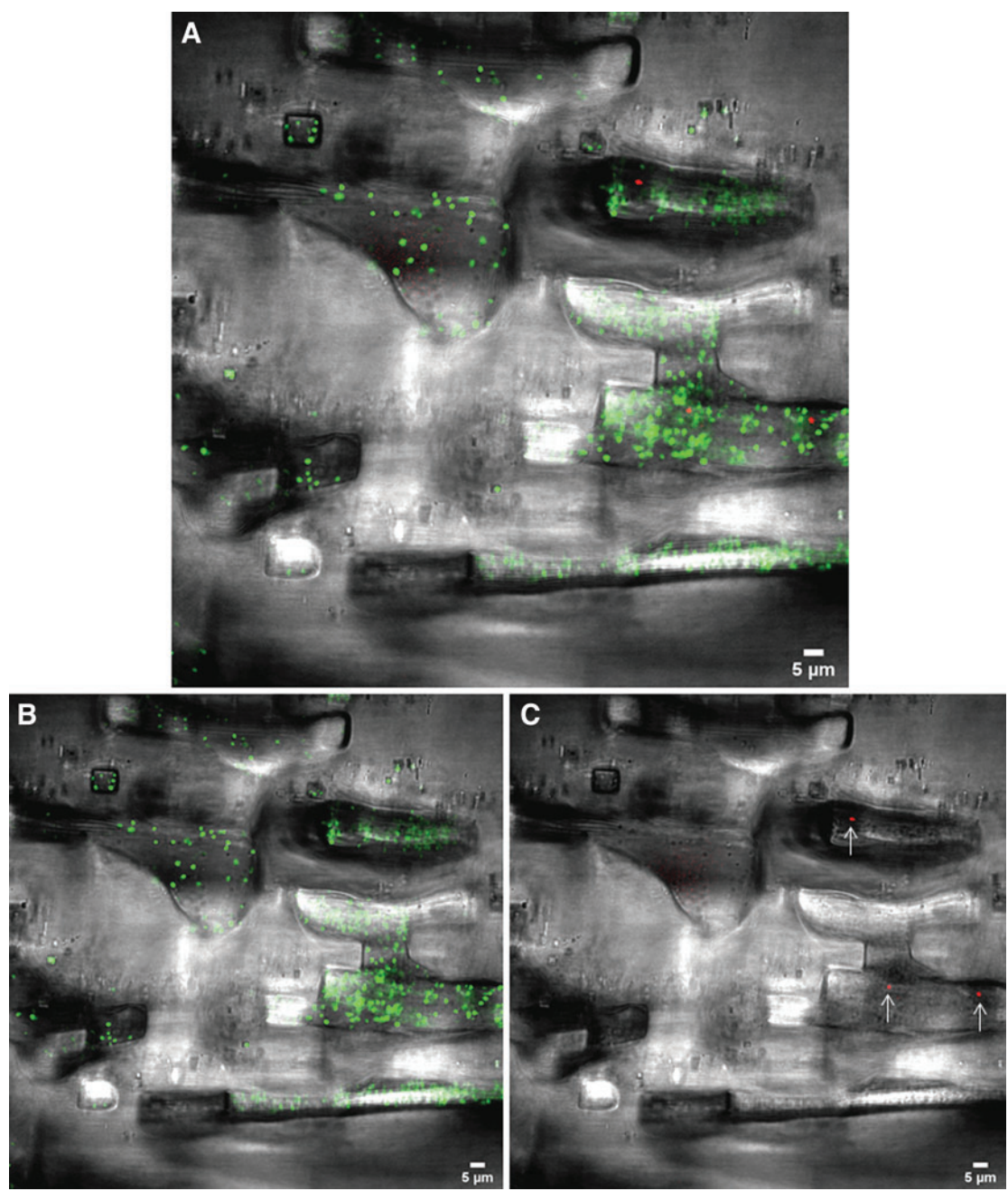

**FIG. 7.** Confocal fluorescence microscopy images (1000×) of salt crystal-embedded cells of *Haloferax volcanii* co-stained with SYTO 9 and PI, showing characteristics of the crystal, the distribution of microorganisms inside fluid inclusions, and potential damage to the cell membrane. **(A)** Color composite (SYTO 9 and PI). **(B, C)** SYTO 9 and PI signal, respectively. The arrows indicate cells with PI signal without SYTO 9 signal.

The results for salt crystal-embedded cells show an exponential decrease up to a fluence of 2000 J/m², which corresponds to superflare VUV fluences for all considered planetary orbits. The results for halite crystal-embedded cells also show an exponential decrease in survival up to a fluence of 2500 J/m², which also corresponds to superflares as in the previous case (Fig. 10).

The analysis of the initial slopes of the curves (*k*-value, Table 3) shows a decrease in the *k* value for *H. volcanii* in halite in comparison with the *k*-value for *H. volcanii* in saline solution, which can be attributed to the enhanced protective effects of the halite to VUV. This can be visualized through the Ho$_{90}$ (Table 3) value, which shows that it would be necessary to apply 1.5 times more fluence to kill the initial population of *H. volcanii* when the microorganisms are entrapped in the halite crystals in comparison with those that are entrapped in salt crystals.

## 4. Discussion

The goal of our experiments was to mimic the exposure of microorganisms inside halite crystals in meteorites,

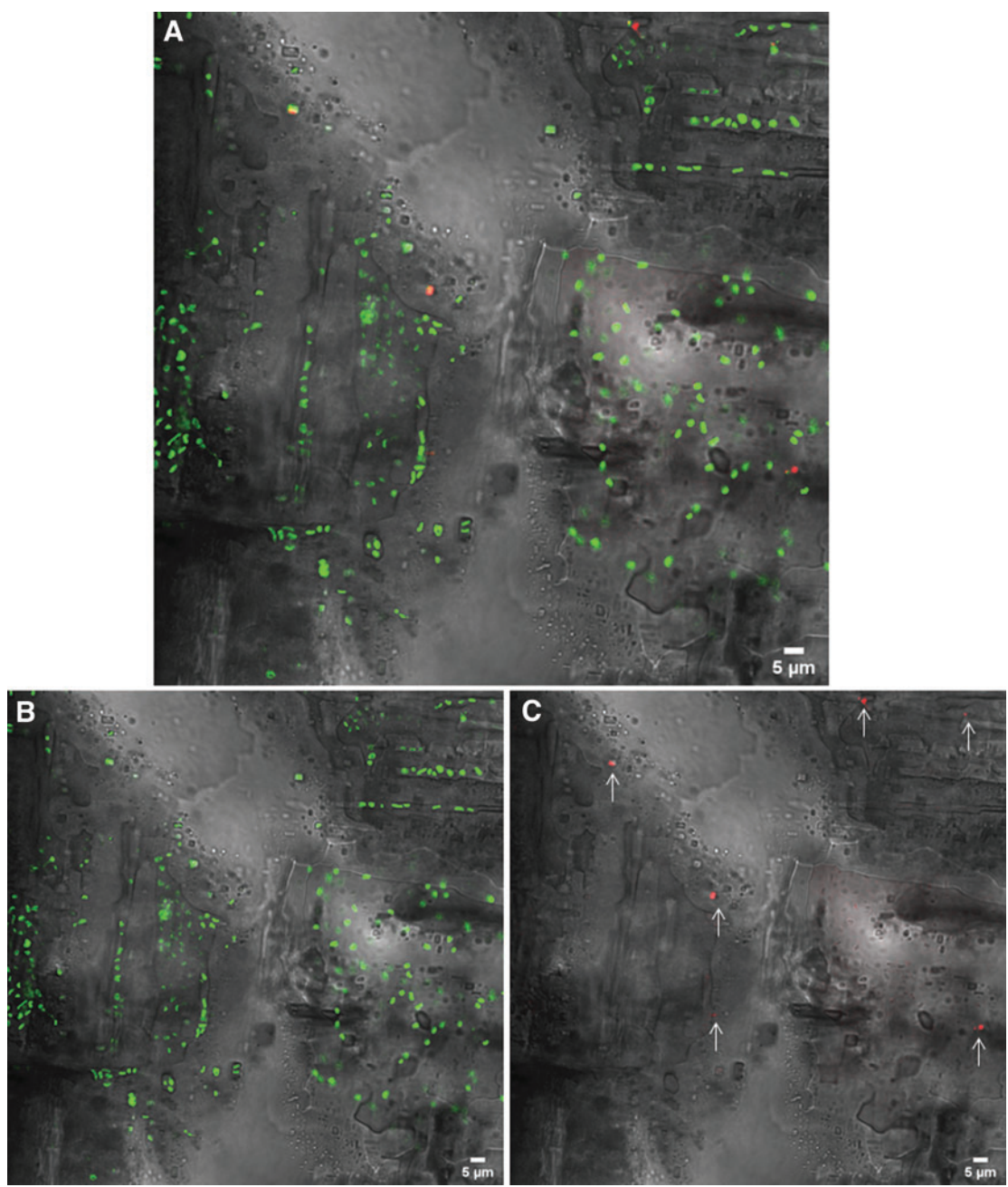

**FIG. 8.** Confocal fluorescence microscopy images (1000×) of halite crystal-embedded cells of *Haloferax volcanii* co-stained with SYTO 9 and PI, showing characteristics of the crystal, the distribution of microorganisms inside fluid inclusions, and potential damage to the cell membrane. **(A)** Color composite (SYTO 9 and PI). **(B, C)** SYTO 9 and PI signal, respectively. The arrows indicate cells with PI signal without SYTO 9 signal.

asteroids, or on the surface of other airless small bodies that lack an atmosphere to some of the detrimental conditions of the interplanetary medium during a lithopanspermia process. Thus, we could evaluate the potential protection that halite might provide to microorganisms embedded in salt crystals exposed to vacuum conditions ($10^{-5}$ Pa) and high VUV fluences (equivalent to those emitted by flares or superflares from a young Sun).

We considered a scenario in the Solar System, 3.8 Gyr ago when the exchange rate of material between planetary bodies was higher than at present and potential habitable planetary bodies in the past where these exchanges could have taken place (Earth, Venus, and Mars).

The entrapment of microorganisms inside salt crystals, such as those composed of NaCl, known as halite, is frequent in natural terrestrial hypersaline environments and might be likely in other planetary bodies such as early Mars (Stan-Lotter *et al.*, 2004). On Earth, halite is formed from the evaporation of surface brines. Fluid inclusions are formed within the crystals as bubbles that may contain micro-samples of the brine (or gas) and are trapped either at the crystal/fluid interface during growth (primary inclusions) or along healed fractures in the mineral (secondary inclusions) (*e.g.,* Roedder, 1984). Fluid inclusions may also contain microorganisms that inhabit brines and become trapped during the evaporation process.

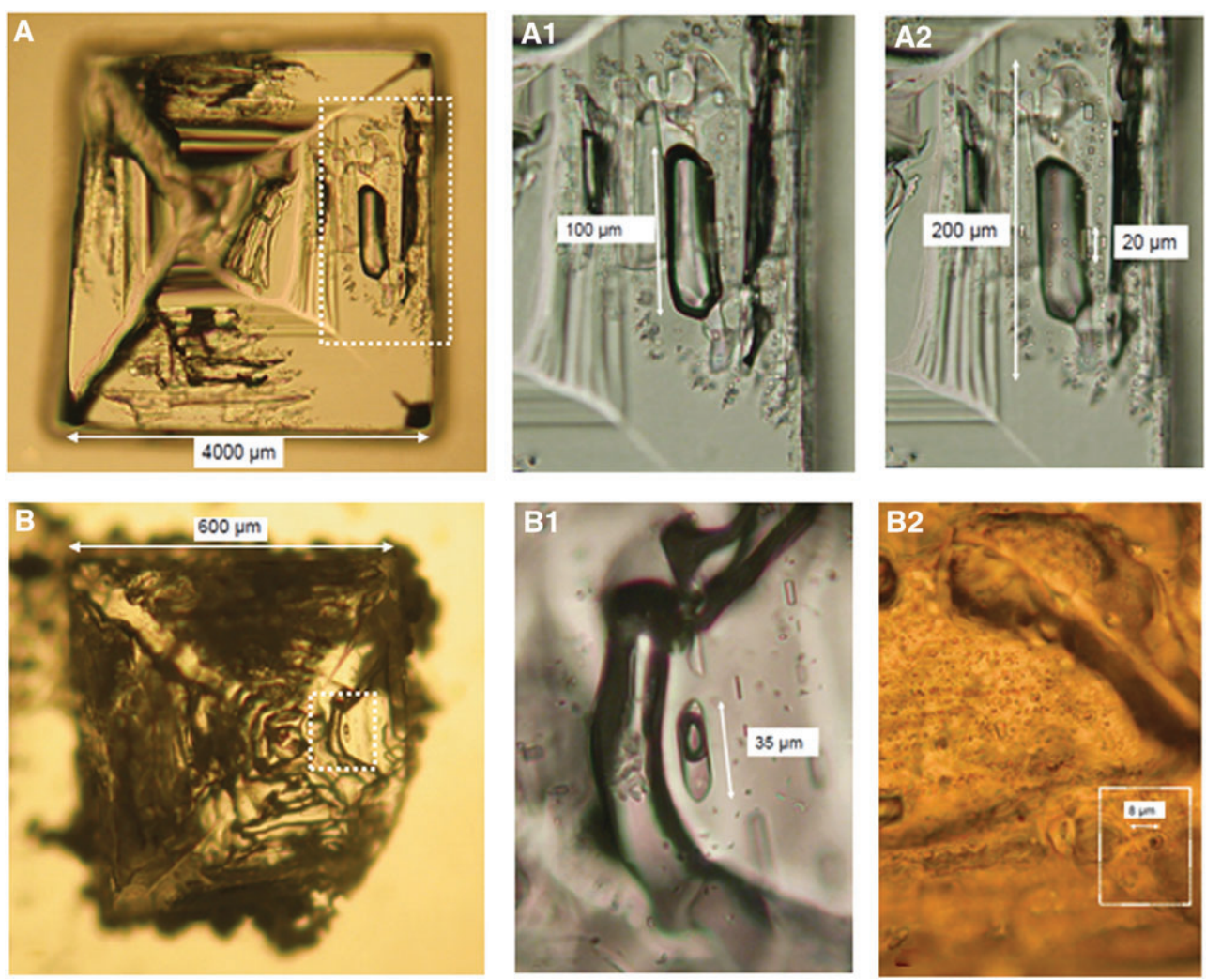

**FIG. 9.** Optical microscopy images showing the morphology and density of fluid inclusions of a halite crystal (**A**) and a salt crystal (**B**), which were obtained in the experiments. (**A1, A2**) Details of fluid inclusions [of the squared section indicated in (**A**)]. (**B1**) Details of fluid inclusions [of the squared section indicated in (**B**)]. (**B2**) Section of the wall of another salt crystal of this sample showing multiple tiny fluid inclusions, and the white square shows a measured fluid inclusion as an example. The bright ''X'' shape observed in halite and salt crystals (**A, B,** respectively) represents the most rapid directions of growth, along which a few inclusions are trapped.

The isolation of microorganisms from ancient halite crystals, such as those from the Permian and Triassic, is plausible evidence of the potential of halite to preserve microorganisms for long periods of time similar to those involved in interplanetary travel associated with lithopanspermia processes (see *e.g.,* Reiser and Tasch, 1960; Dombrowski, 1963; Tasch, 1963; Norton *et al.,* 1993; Grant *et al.,* 1998; Stan-Lotter *et al.,* 2004; Schubert *et al.,* 2010).

Several studies have documented the isolation of microorganisms from ancient rock salt (*e.g.,* Norton *et al.,* 1993; Stan-Lotter *et al.,* 2004; Satterfield *et al.,* 2005; Vreeland *et al.,* 2007; Schubert *et al.,* 2010), which can remain undisturbed for millions of years (Lowenstein *et al.,* 2011;

Schubert *et al.,* 2009a, 2009b). The antiquity of the microorganisms was supported by an independent study (Sankaranarayanan *et al.,* 2014).

Halite is expected to exist on Mars since evaporite mineralogy is apparently widespread on this planet (Moore *et al.,* 2010), with chloride deposits identified across the martian mid-latitudes (Beck *et al.,* 2017). There is evidence of the existence of halite in meteorites, given that the SNC meteorite Nakhla is also considered mineralogical evidence of evaporites on Mars (Bridges and Grady, 1999). Evidence that would support a mechanism of lithopanspermia when considering halite crystals comes from the fact that halite minerals have been found on Mars and in meteorites (Bridges and Grady, 1999; Gooding *et al.,* 1991).

Thus, our study of lithopanspermia, which was driven by implementation of halite crystals, is well motivated. Further motivation has to do with halite as a shielding mechanism against radiation and the potential for long-term survival of microorganisms in fluid inclusions that enhance and support the idea of a successful interplanetary journey. Before the present study, microorganisms had only been exposed to conditions that do not properly evaluate the effects of VUV and vacuum.

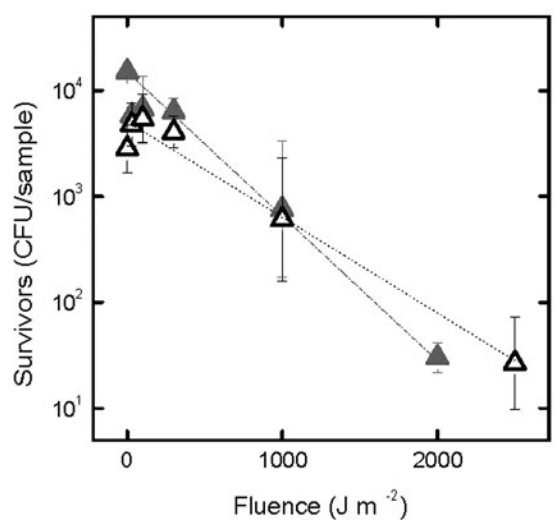

**FIG. 10.** *H. volcanii* salt crystal-embedded cells (gray triangles) and halite crystal-embedded cells (white triangles) exposed to different fluences of VUV plus vacuum ($N_R$). Vertical bars indicate standard errors. Dotted lines represent the adjustment to non-linear regression.

To mimic the interplanetary step of lithopanspermia when considering halite crystals in meteorites as potential carriers of life, we exposed both non-halophilic and halophilic microorganisms that were entrapped in halite. The conditions that were tested to evaluate the protective effects of halite were vacuum and VUV. The results of cells embedded in halite crystals were compared with those where the microorganisms were not entrapped inside halite.

### 4.1. Effects of air-drying and hyperosmotic conditions (halite entrapment)

As we exposed the microorganisms to vacuum and VUV after water evaporation, the starting point of our experiments was to evaluate the effects of air-drying on naked cells (*D. radiodurans*) and halite/salt crystals (*D. radiodurans* and *H. volcanii*), respectively. We analyzed potential changes in cell morphology and distribution (as clumping, for instance, may shield UV) and potential damage to cell membranes given that previous studies documented that this might be the main source of VUV damage (see Section 4.3).

A co-staining experiment with SYTO 9 and PI revealed that, in the case of naked cells of *D. radiodurans*, the cellular membranes were only partially damaged, and the microorganisms appeared clumped and grouped in monolayers. This is a kind of typical cell distribution that has already been observed for this microorganism under optimum growth conditions (*e.g.*, Slade and Radman, 2011).

In the case of halite crystal-embedded cells of *D. radiodurans*, the microorganisms remained distributed inside fluid inclusions and did not show evidence of clumping. No substantial cell membrane damage was found in the co-staining experiments. This is at odds with a previous study that indicated that salts can increase multi-cell formations in *D. radiodurans*, although this particular study was conducted such that concentrations of up to 1% NaCl were added to the culture medium, which represents a concentration of NaCl 20 times smaller than that used in the present study (Chou and Tan, 1991).

According to our results, a considerable fraction of *D. radiodurans* cells could remain viable inside fluid inclusions of laboratory-grown halite (at least for the duration of our experiments of around 24–48 h). Given that liquid fluid inclusions contain the same brine that originated the crystal, our results show that *D. radiodurans* can withstand hypersaline/hyperosmotic conditions.

It should be noted that a study by Im et al (2013) supports our findings of potential osmotolerance features in *D. radiodurans*. These authors identified genetic expression associated with osmoprotectant accumulation in this microorganism during salt stress induced by NaCl (particularly, an operon encoding a high-affinity phosphate transporter, and genes encoding components of transporters of glycine betaine and trehalose).

A previous study in which *D. radiodurans* was exposed to 25% NaCl in the culture medium for 2 months failed to detect survivors (Mancinelli *et al.*, 2004). This study discarded osmotolerance (osmophilic) features in this microorganism, but this result was potentially influenced by the physiological state of the cells, which were in the mid-log phase culture (instead of a late exponential culture as in the case of our experiments) or even by the particular experimental conditions during that assay, which included five washings by centrifugation and could have induced cell injuries.

Although our study is the first article that shows the survival of *D. radiodurans* inside fluid inclusions of laboratory-grown halite, a previous study indicated that halite crystal-embedded cells of the non-halophilic bacterium *Pseudomonas aeruginosa* could survive inside fluid inclusions for at least 6 weeks (Adamski *et al.*, 2006).

For the case of the entrapment of the *H. volcanii* in salt or halite crystals, the co-staining experiment has shown that, as for *D. radiodurans*, the cells were located inside fluid inclusions (not in the crystal matrix). No damage to the cell membrane has been detected for both salt and halite crystal-embedded *H. volcanii*. This result is completely expected, as these processes of halite entrapment occur in natural hypersaline environments, and haloarchaea are adapted to thrive within a wide range of high salt concentrations and embedment in salt or halite crystals. Entrapment in laboratory-grown halite was already documented for haloarchaea (*e.g.*, Fendrihan *et al.*, 2009).

Interestingly, we have identified different cell morphologies inside fluid inclusions for *H. volcanii*, depending on whether the entrapment was made in salt crystals (a mixture of different salts containing 2.5 *M* NaCl) or in halite crystals (containing only 4 *M* NaCl). The co-staining experiments have revealed that *H. volcanii* adopted rod-cell shapes inside fluid inclusions of salt crystals and coccoid-cell shapes when entrapped inside fluid inclusions of halite crystals.

It has already been shown that *H. volcanii* is a pleomorphic microorganism, a phenomenon that exists in bacteria but is especially common in haloarchaea, in which a cellular population changes its morphology depending on environmental factors or growth phase (Oren, 2014; Schwarzer *et al.*, 2021). The great morphological plasticity that has been described for *H. volcanii* reveals cellular changes in shape and size, as discoid-shaped forms, rods, motile rods, and long filaments.

This can occur in response to multiple factors, including changes in medium composition (De Silva *et al.*, 2021; Schwarzer *et al.*, 2021). Fendrihan *et al.* (2012) described an additional phenomenon of the formation of ''small roundish'' or spherical particles in haloarchaea of *Halobacterium* species when exposed to low water activity. The same kind of particles have been found in fluid inclusions of 22,000- to 34,000-year-old halite and identified as haloarchaea capable of proliferation (Schubert et al, 2010). These morphologies are similar to those found in fluid inclusions of halite crystals in our experiments for the case of *H. volcanii*, and we could not discard a similar phenomenon, although more experiments would be necessary to confirm this fact.

Therefore, cells entrapped inside salt crystals and exposed to hyperosmotic conditions did not show evidence of substantial membrane damage, from both halophiles and nonhalophiles. Exposure to air drying without the protection of the crystal showed partial damage to the membrane, as the experiment with naked cells of *D. radiodurans* has evidenced.

### 4.2. Effects of vacuum

After the air-drying step, the microorganisms were introduced into a vacuum chamber, and they were kept under these conditions before, and during, all the subsequent irradiation experiments. We simulated vacuum conditions of the space environment, particularly at values that are possible to find at low Earth orbit ($10^{-4}$ Pa). The total exposure time to vacuum during the experiments was around 15 h.

Our results show that there was a decrease of four orders of magnitude in the survival of naked cells of *D. radiodurans* when exposed to vacuum. This is in agreement with a previous study that indicated that exposure to the space vacuum decreased cell survival by four orders of magnitude for *D. radiodurans*, even higher than for other bacteria such as *Bacillus* sp. strain PS3D, which showed a decrease in cell survival of two orders of magnitude (Saffary et al., 2002). In the case of halite-embedded cells of *D. radiodurans*, the decrease in survival was between two and three orders of magnitude.

When comparing survival fractions with vacuum, haliteembedded cells have values of one order of magnitude higher than naked cells, which shows that halite crystals are effective in protecting the cells against the effects of vacuum.

Similar results were obtained for *H. volcanii*. There was a decrease of three orders of magnitude in the survival for salt crystal-embedded cells; whereas, when exposed to vacuum in halite crystals, the decrease was of two orders of magnitude. Interestingly, this comparison also shows that halite crystals seem to offer better protection to vacuum than that provided by the salt crystal formed by the evaporation of a mixture of salts. This could be related to the different geometries that salt crystals can develop according to the mineral composition. In the case of 4 *M* NaCl, the crystal featured an expected cubic geometry (Fig. 8).

In the case of saline solution, the geometry of the crystal, due to the mixed salt composition, seems to be a combination of different geometries such as cubic (*e.g.*, NaCl+KCl), orthorhombic (*e.g.*, $MgSO_4.7H_2O$; $CaCl_2$), and monoclinic (*e.g.*, $MgCl_2.6H_2O$) (Fig. 7).

The results obtained in our experiments regarding the protection of halite against vacuum can be explained by the fact that halite-embedded cells (and specifically in the liquid brine of fluid inclusions) can avoid the effects of vacuum that causes almost complete dehydration of microbial cells through water desorption. This avoids damage to the cell membrane, among other components, that severely affects the integrity of macromolecular structures (Horneck, 1981).

For instance, damage at the level of DNA was documented at a humidity below 65%, as DNA changes its configuration and assumes an A form, which therefore affects its photochemical and photobiological properties (Horneck, 1981). Other experiments in laboratory conditions have shown that cells resuspended after vacuum treatment show membrane leakage and loss of osmotic control, apart from being incapable of membrane-mediated cell wall control biosynthesis.

However, during subsequent incubation, a fraction of the microbial population was able to repair the membrane damage (which can be seen as a delay during the lag phase) and resume growth (Horneck, 1981).

The protection offered by halite crystals is related to findings of previous reports that have shown that the inactivation of microorganisms by vacuum treatment alone depends on the exposure conditions. Partial protection was achieved by cell stacking in multicellular layers or by the addition of glucose, salts, or nutrient broth. Even though ultra-high vacuum alone (at the level found in space conditions) inactivates cells, this would not prevent the survival of microorganisms in both the form of spores and vegetative forms (Portner *et al.*, 1961; Horneck, 1981; Koike *et al.*, 1992).

For instance, it has been shown that, in general, around 2–5% of the vegetative cell population survives without any protection during 20 days of exposure to vacuum ($2.66 \times 10^{-7}$ Pa), whereas spores of *Bacillus subtillis* completely survived the vacuum treatment (Horneck and Bücker, 1980). In our experiments, the fraction of cell survivors was lower. For non-embedded cells of *D. radiodurans and H. volcanii* in halite, the percentage of survival to vacuum was 0.017% (naked cells) and 0.37% (salt crystal-embedded cells), respectively, whereas for embedded cells of *D. radiodurans* and *H. volcanii* in halite it was 0.83% and 1.25%, respectively.

Therefore, the increase in the survival observed for microorganisms entrapped inside fluid inclusions in halite crystals exposed to vacuum can account for the avoidance of extreme dehydration and its consequent cellular damage.

### 4.3. Effects of vacuum and VUV from stellar flares and superflares

As young stars are known to be very active, we simulated flare or superflare emissions (fluences) in the VUV range. Neither this effect nor the potentially protective effects of halite had been properly evaluated in previous studies as there are little data from biological experiments using VUV radiation below 124 nm, which justified our choice of the wavelength range 57.6–124 nm. Previous studies have revealed that simultaneous exposure to space vacuum and VUV radiation in laboratory simulations (at lower VUV fluxes than in our experiments) decreased the survival of microorganisms compared with unirradiated control samples (see *e.g.*, Horneck and Bücker, 1980; Saffary *et al.*, 2002; Paulino-Lima *et al.*, 2010; Abrevaya *et al.*, 2011; Araujo *et al.*, 2020).

For the estimation of the VUV fluxes, we used the solar analog $\kappa^1$ Cet. We considered a worst-case scenario where life-forms would be completely exposed to the solar UV radiation during a superflare event inside halite in a meteorite, disregarding that meteorites and planetary bodies orbit around the Sun and the incident UV flux at any point on the surface can vary due to a number of factors (*e.g.*, orbital distance, Sun elevation angle, latitudinal changes, rotational periods, positions on the "day" or "night" sides of objects, and shading among others) (Nicholson *et al.*, 2005).

We analyzed several cases considering the possibility that the meteorite could be close to or in different planetary orbits (Venus, Earth, Mars). As expected, different flux enhancement factors with respect to the quiescent level of flares or superflares would be required for different orbital distances to reach fluxes and fluences that are equivalent to the VUV flux and fluences used in our experiments.

The exposure of halophilic and non-halophilic microorganisms to our experimental conditions with a VUV flux of 7.14 $W/m^2$ and a pressure of $10^{-4}$ Pa showed different survival responses depending on the assay condition (naked cells, halite-embedded cells, and salt crystal-embedded cells) and microorganisms under consideration. Nevertheless, for all cases, our results demonstrate that a part of the population of microorganisms would be able to withstand the VUV radiation at fluences that they would receive from flares or superflares in different planetary orbits (Venus, Earth, Mars), and that the embedment within halite crystals significantly increases the chances of survival, as seen by the change in the slope of the survival curves (Figs. 6 and 10).

For the particular case of naked cells of *D. radiodurans*, the survival curve showed a biphasic behavior. There was a rapid decrease in survival at the beginning (up to fluences of 300 $J/m^2$) and a slower decrease seen as a tail, which can be interpreted as the survival of a small fraction of the population. This kind of survival curve has been obtained in previous studies in which the same microorganism were exposed to VUV and vacuum (Paulino-Lima *et al.*, 2010; Abrevaya *et al.*, 2011; Araujo *et al.*, 2020).

In previous studies, the survival of a small fraction of the population seen as a tail in the survival curve was accredited to shielding, given the characteristics of the matrix where the cells were deposited and irradiated (Paulino-Lima *et al.*, 2010) or shielding produced by cell clumping or salts (Araujo *et al.*, 2020). However, the biphasic survival curve seen in our experiments (Fig. 6) was obtained in the absence of any source of shielding. This shows that an alternative explanation other than shielding would be necessary to understand these results. We argue that this could be based on the existence of different phenotypes inside an exposed microbial population (Pennell *et al.*, 2008).

In contrast, the results of *D. radiodurans* entrapped in halite show an exponential decrease in survival with a lower $k$-value and higher $Ho_{90}$ value than for naked cells. Although this curve does not show a change in the slope in the fluence range assayed, a decrease in the slope could eventually happen at higher fluences. Such a change would be undetectable because the expected survival fractions would be below the LOD. The survival to higher fluences should not be predicted from the results obtained at low fluences (Abrevaya *et al.*, 2018, 2020).

A comparison of the $Ho_{90}$ values for naked entrapped cells of *D. radiodurans* and those of halite shows that 11 times more VUV fluence is needed to kill 90% of the population of microorganisms when *D. radiodurans* is embedded in halite with respect to naked cells. The protective effects of halite to VUV can be attributed to the shielding provided by the crystal, which could prevent damage to the cell membrane and its surface structures classically attributed to radiation at these levels (Coohill, 1986; Ito, 1989).

However, other mechanisms may be operating, as will be described later, which are related to the formation of free radicals and increased UV sensitivity induced by vacuum. For instance, in a previous experiment with *D. radiodurans* cells (dried in 0.9 g·L NaCl) that had been exposed to VUV synchrotron radiation in the same wavelength range as our experiments, it was shown that VUV inactivating effects can be extended beyond the cell membrane, as the membrane damage was increased at a much lower rate than the cellular inactivation (Araujo *et al.*, 2020).

As intracellular oxidative stress was detected on exposure of the cells to VUV, and only a fraction of the cells presented membrane damage, the production of free radicals could be the main damaging mechanism that explains these results (Araujo *et al.*, 2020). An alternative hypothesis proposed in the same study discusses the possibility of endogenous reactive oxygen species (ROS) that could be generated as part of the experimental process when cells are rehydrated for recovery.

This would be increased by cells with membrane damage that try to restart their metabolism ineffectively, as the respiratory chain of aerobic organisms is known to generate these types of compounds (Cabiscol *et al.*, 2010). Nevertheless, there is evidence about the formation of ROS from experiments using plasmid DNA that have shown single-strand breaks by an indirect mechanism that involves the production of OH-radicals through direct photolysis of water induced by VUV (with a maximum at 150 nm) (Takakura *et al.*, 1986).

The formation of these radicals is in agreement with the fact that, for the VUV region, higher excitation, ionization, and dissociation of molecules will occur, whereas for the far-UV region lower electronic excitation prevails (Wehner and Horneck, 1995). DNA damage is also expected. For instance, previous studies have shown that, in the UV region below 180 nm, the chain moiety of DNA (deoxyribose-5′phosphate) is the main target of the photons (Hidea *et al.*, 1986) and wavelength-dependent formation of photoproducts is obtained for DNA molecules in UV ranges between 150 and 365 nm, including base changes such as cyclobutane dipyrimidines and (6-4) pyrimidine-pyrimidone (Matsunaga *et al.*, 1991; Yamada and Hieda, 1992).

Moreover, VUV photolysis was also found (for wavelengths <200 nm), which leads to the rupture of the N-glycosidic bond with the liberation of bases. In opposition to the photo-destruction of nucleotides by VUV, no bases have been found among the products of near UV photolysis (Dodonova *et al.*, 1982). Single-strand breaks have also been identified, probably related to the mechanism of the liberation of bases described earlier (Hidea *et al.*, 1986; Sontag and Dertinger, 1975; Wirths and Jung, 1972).

It is important to note that, as this wavelength of UV radiation requires vacuum to be propagated, the exposure to VUV implies simultaneous exposure to vacuum, and we cannot separate the effect of radiation from vacuum, since we envisage a situation in the interplanetary space. This is important for two reasons. First, in the case of naked cells, although a decrease in the potential to generate free radicals on VUV irradiation due to a decrease of intracellular water under ultra-high vacuum (and therefore in presence of water loss by desorption) would be expected, it is not possible to totally exclude the influence of water radicals (Wehner and Horneck, 1995).

For example, Swarts *et al.* (1992) showed that tightly bound water molecules of the DNA hydration layer contribute to radiation-induced DNA damage primarily via the transfer of charges from these water molecules into the DNA, a mechanism that is different from the induction of DNA damage after irradiation of bulk water [OH.,e(aq-)]. Moreover, Saenger (1984) suggested that water molecules bind predominantly to the sugar-phosphate group at low levels of hydration (probably even in vacuum) and, with increasing levels of hydration, also to the bases, allowing the formation of water radicals. It has also been documented that DNA bases and ribose-phosphate can be ionized by VUV photons in aqueous solutions (Candeias and Steenken, 1992).

A second reason why it is important to consider the effect of vacuum is that it can increase the sensitivity to UV radiation (Horneck and Bücker, 1980). As described in Section 4.2, vacuum can induce structural changes in the DNA, including change of the structural conformation and partial denaturation, and probably in proteins, thus creating synergistic effects between vacuum and VUV radiation (Horneck, 1981). For instance, a tenfold increase in UV sensitivity was obtained for *B. subtilis* spores exposed to the space vacuum, compared with spores irradiated at atmospheric pressure (Horneck, 1998).

Apart from these structural changes induced to DNA by vacuum, it was found that specific photoproducts were formed under vacuum conditions, such as DNA-protein cross-links, 5,6-Dihydro-5(α-thyminyl)thymine, and trans-syn thymine dimer. At least one of these lesions may be responsible for the increased UV sensitivity in vacuum as these are lesions that are not, or slightly less, repairable by the cell enzymatic machinery (Horneck, 1981). In particular, this has been ascribed to trans-syn thymine dimer because it requires at least partially denatured DNA and can be attributed to vacuum (Horneck, 1981).

The increased survival seen for microorganisms entrapped in halite crystals can be related to the fact that halite can create an environment where microorganisms can avoid vacuum, as they are entrapped inside fluid inclusions that could also help to reduce the damaging effects of VUV by reducing the UV sensitivity.

The experiments made with salt crystal and halite crystal-embedded cells of *H. volcanii* provided additional information. The Ho90 value for microorganisms entrapped in salt crystals was around $733 \, J/m^2$, which would be the flu-ence necessary to kill 90% of the population of microorganisms. The Ho90 value obtained for microorganisms entrapped in halite was higher, around $1001 \, J/m^2$.

The comparison between both Ho90 values reveals that, although in both cases the cells were shielded inside fluid inclusions, this reference value is 1.5 times higher for halite crystal-embedded cells than for salt crystal-embedded cells, and therefore, the protection given by halite crystals was higher than that provided by salt crystals. Surprisingly, crystals seem not to be merely a physical barrier that protects from the effects of VUV, but also that protection has to do with its mineral composition.

A previous study using UV radiation in the 200–400 nm range showed that halite crystals can absorb and re-emit UV radiation at longer wavelengths, due to the absorption of the color centers in halite (Fendrihan *et al.*, 2009). Although we do not have data about the absorption spectrum characteristics of halite and salt crystals in the VUV range, a potential explanation as to why salt crystals offer less protection to VUV than halite crystals could be related to their different morphologies, because of their specific mineral composition.

As described in Section 4.2., salt crystals are characterized by developing a ''hopper'' shape that will increase the presence of defects in the structure of the crystal (*e.g.*, the presence of clusters of lattice vacancies). This could explain the variation in the related absorption bands of incoming light in ionic crystals due to color centers. In addition, we cannot rule out that reflectance and transmittance of UV could have varied due to the potential impurities that may have been present together with $MgSO_4.7H_2O$ (such as Ni, Fe, Co, Mn), consequently affecting the effectiveness of shielding against VUV radiation.

Moreover, as halite crystals seem to provide better protection from vacuum than salt crystals (Table 2), it would be expected that the synergistic effects between vacuum and VUV are decreased in microorganisms trapped inside halite crystals when compared with those in salt crystals.

From the Ho90 values (Table 3), we could infer that the emission of VUV from a flare would be enough to kill only 90% of the population of microorganisms at the orbits of Venus, Earth, or Mars, considering as a model naked cells of *D. radiodurans*.

In the case of microorganisms entrapped in halite crystals or salt crystals, the chances of survival increase. For *H. volcanii* entrapped in salt crystals, the VUV emitted to kill 90% of the population would be in the range of flares or superflares depending on the planetary orbit, which is similar to the case of naked cells of *D. radiodurans*. For *D. radiodurans* and *H. volcanii* entrapped in halite, VUV fluences such as those that come from superflares would be necessary to kill 90% of the population, for all considered planetary orbits, except for *H. volcanii* in the orbit of Venus, which would correspond to the intensity of a normal flare.

However, it is important to emphasize that the Ho90 values should be used for comparative reasons to evaluate the protective effects of halite and must not be taken as predictors of the survival of the population of microorganisms when exposed to flares or superflares because of the potential existence of biphasic survival curves (Abrevaya *et al.*, 2020).

Considering the traveling times of meteorites between planetary bodies of the Solar System, estimated in approximately millions of years, although our experiments do not offer enough data to mimic the effect of the exposure to repetitive flares, the latter factor should be taken into consideration. The long-term survival would be influenced by the capability of growth of the population after a flare. It is unknown how non-sporulating microorganisms such as haloarchaea can survive for long periods of time inside halite fluid inclusions.

If the microorganisms enter a dormant state, cell division would be blocked, and DNA damage produced by successive irradiation events could accumulate. Then, the effects of radiation on the overall population can be more deleterious, decreasing the chances of survival, especially over long periods of time (million years). Conversely, if cell division after flare events is possible, then the chances of survival would be increased, as the mechanisms of DNA repair would be active, and the survivors of the irradiated population could proliferate and restore the original population.

Our estimations show that the microorganisms would be subjected to superflares at a rate of $\sim 70$ events per year. Nevertheless, this exposure rate would be significantly lower, as some orbital models predict that, in the case of some martian meteorites, the traveling time could last a few years only, particularly for those that can reach Earth-crossing orbits on a fast track (Gladman, 1997).

Moreover, the microorganisms would not be necessarily exposed to every event, as it is necessary to consider the position of the rocky body relative to the position of the flare on the Sun and the rotation of this body (*e.g.*, meteorite). In this case, the avoidance of exposure to radiation—but, of course, not to vacuum—would increase the chances of survival.

It is important to stress that the VUV flux used in our experiments (equivalent to a superflare emitted by the young Sun) is several orders of magnitude higher than that which microorganisms would receive in an open space experiment in the low Earth orbit with radiation coming with the present Sun in quiescence. As stated in Section 3.1, the quiescent VUV flux of $\kappa^1$ Cet is 1–2 orders of magnitude higher than that of the present Sun, and the TGM flux is a factor of 155 higher than the quiescent VUV flux of $\kappa^1$ Cet.

Therefore, the VUV flux used in our experiments is about 3–4 orders of magnitude above the flux of the present Sun. Following this reasoning, the microorganisms in our experiments received VUV fluxes representative of superflares 3–4 orders of magnitude higher than those that microorganisms would receive in experiments performed in space in low Earth orbit. In this sense, our study is a new contribution in that space experiments done in low Earth orbit are not representative of conditions of a superflare scenario from the young Sun nor would extrapolations from space experiments be accurate as the biological impact would be different.

In agreement with our results, we hypothesize that the protective effects of halite probably come from the prevention of the excitation of molecules generated by VUV photons, which produce cellular damage, and by prevention of the synergistic effects of VUV with the vacuum, which increase the lethal effect of this type of radiation.

## 5. Conclusions

Our experiments have shown, for the first time, that halite crystals provide protection to vacuum (at the levels of low Earth orbit) as well as to high VUV fluxes (*i.e.*, radiation levels related to flares and superflares in our assays covering the range 57.6–124 nm). This protective effect is seen regardless of the type of microorganism. We have demonstrated that the protection by halite crystals would not be limited to haloarchaea but can be potentially extended to non-halophilic bacteria such as *D. radiodurans*. In addition, our study shows that this bacterium can withstand the physiological effects of entrapment inside halite, exhibiting considerable viability and resilience.

Moreover, it is important to emphasize the high protection that halite provides in comparison with crystals made of salt mixtures (even when they contain NaCl among other minerals in their composition).

These findings contribute to extending our understanding of the protection of halite as documented in previous studies that considered only the 200–400 nm UV range. Moreover, our results agree with the fact that halite crystals may be structures that protect microorganisms against a set of factors (in particular VUV radiation and vacuum) and, therefore, would significantly increase the chances of survival of microorganisms in airless bodies, for instance, in the context of lithopanspermia.

## Acknowledgments

The authors are grateful for the invaluable technical support of Evandro da Silva from *Núcleo de Pesquisas em Astrobiología*, Brazil, during the experiments. This research used resources from the LNLS, an open national facility operated by the CNPEM for the Brazilian Ministry for Science, Technology, Innovations, and Communications (MCTIC). The TGM beamline staff is acknowledged for their assistance during the experiments.

## Author Disclosure Statement

No competing financial interests exist.

## Funding Information

X.C.A. acknowledges CNPEM for the beamtime granted to the proposal TGM—16126 (LNLS), FAPESP postdoctoral fellowship (years 2013–2014) (Processo nro: 2012/20106-5), Brazil, and funding from PIP—CONICET 0754, Argentina. M.L. and P.O. acknowledge the Austrian Science Fund (FWF): P30949-N36, I5711-N for supporting this project. J.E.H. acknowledges the financial support of FA-PESP (São Paulo State) and CNPQ (Brazil) financing agencies. G.F.P.M. acknowledges grant 474972/2009-7 from CNPq/Brazil.

## References

Abramov O, Mojzsis SJ. Microbial habitability of the Hadean Earth during the late heavy bombardment. *Nature* 2009; 459(7245):419–422.

Abrevaya XC, Hanslmeier A, Leitzinger M, *et al.* The BIOSUN project: An astrobiological approach to study the origin of life. *Rev Mex Astron Astrofis Conf Ser* 2014;44:144–145.

Abrevaya XC, Leitzinger M, Oppezzo OJ, *et al.* Towards astrobiological experimental approaches to study planetary UV surface environments. International Astronomical Union. *Proc Int Astron Union* 2018;14(S345):222–226.

Abrevaya XC, Leitzinger M, Oppezzo OJ, *et al.* The UV surface habitability of Proxima b: First experiments revealing probable life survival to stellar flares. *MNRAS Lett* 2020;494(1):L69–L74.

Abrevaya XC, Paulino-Lima IG, Galante D, *et al.* Comparative survival analysis of *Deinococcus radiodurans* and the haloarchaea *Natrialba magadii* and *Haloferax volcanii* exposed to vacuum ultraviolet irradiation. *Astrobiology* 2011;11(10):1034–1040.

Adamski JC, Roberts JA, Goldstein RH. Entrapment of bacteria in fluid inclusions in laboratory-grown halite. *Astrobiology* 2006;6(4):552–562.

Anderson AW. Studies on a radioresistant micrococcus. 1. Isolation, morphology, cultural characteristics, and resistance to γ radiation. *Food Technol* 1956;10:575–578.

Antunes A, Olsson-Francis K, McGenity TJ. Exploring deep-sea brines as potential terrestrial analogues of oceans in the icy moons of the outer Solar System. *Curr Issues Mol Biol* 2020;38:123–162.

Araujo GG, Rodrigues F, Galante D. Probing the response of *Deinococcus radiodurans* exposed to simulated space conditions. *Int J Astrobiol* 2020;19:203–209.

Arrhenius S. The spread of life in Space. The overview. [Die Verbreitungdes Lebens im Weltenraum.] *Die Umschau* 1903; 7:481–485.

Audard M, Güdel M, Drake JJ, *et al.* Extreme-ultraviolet flare activity in late-type stars. *ApJ* 2000;541:396–409.

Balona LA. Flare stars across the H-R diagram. *MNRAS* 2015; 447:2714–2725.

Barber DJ. Matrix phyllosilicates and associated minerals in C2M carbonaceous chondrites. *Geochim Cosmochim Acta* 1981;45(6):945–970.

Battista JR. Against all odds: The survival strategies of *Deinococcus radiodurans. Annu Rev Microbiol* 1997;51(1):203–224.

Bauermeister A, Moeller R, Reitz G, *et al.* Effect of relative humidity on *Deinococcus radiodurans*' resistance to prolonged desiccation, heat, ionizing, germicidal, and environmentally relevant UV radiation. *Microb Ecol* 2011;61(3): 715–722.

Beck AW, Viviano-Beck CE, Murchie SL, *et al.* CRISM mapping of chlorides deposits at central latitudes in Mars-Noachis Terra. Abstract No. 2247, *Lunar and Planetary Science XLVIII,* Lunar and Planetary Institute: Houston; 2017.

Bell EA, Boehnke P, Harrison TM, *et al.* Potentially biogenic carbon preserved in a 4.1 billion-year-old zircon. *Proc Natl Acad Sci U S A* 2015;112:14518–14521.

Berkley JL, Taylor GJ, Keil K, *et al.* Fluorescent accessory phases in the carbonaceous matrix of ureilites. *Geophys Res Lett* 1978;5(12):1075–1078.

Boulos L, Prevost M, Barbeau B, et al. LIVE/DEAD® BacLight™: Application of a new rapid staining method for direct enumeration of viable and total bacteria in drinking water. *J Microbiol Methods* 1999;37(1):77–86.

Braslavsky S. Glossary of terms used in photochemistry, 3rd edition (IUPAC recommendations 2006). *Pure Appl Chem* 2007;79:293–465.

Bridges JC, Grady MM. A halite-siderite-anhydrite-chloroapatite assemblage in Nakhla: Mineralogical evidence for evaporites on Mars. *Meteorit Planet Sci* 1999;34(3):407–415.

Cabiscol E, Tamarit J, Ros J. Oxidative stress in bacteria and protein damage by reactive oxygen species. *Int Microbiol* 2010;3:3–8.

Candeias LP, Steenken S. Ionization of purine nucleosides and nucleotides and their components by 193-nm laser photolysis in aqueous solution: Model studies for oxidative damage of DNA. *JACS* 1992;114(2):699–704.

Chou FI, Tan ST. Salt-mediated multicell formation in *Deinococcus radiodurans. J Bacteriol* 1991;173(10):3184–3190.

Coohill TP. Virus-cell interactions as probes for vacuum-ultraviolet radiation damage and repair. *Photochem Photobiol* 1986;44(3):359–363.

Davenport JRA. The Kepler catalog of stellar flares. *ApJ* 2016; 829:23.

De Silva RT, Abdul-Halim MF, Pittrich DA, *et al.* Improved growth and morphological plasticity of *Haloferax volcanii. Microbiology (Reading)* 2021;167(2):001012.

Demets R, Schulte W, Baglioni P. The past, present and future of Biopan. *Adv Space Res* 2005;36(2):311–316.

Dodonova NY, Kiseleva MN, Remisova LA, *et al.* The vacuum ultraviolet photochemistry of nucleotides. *Photochem Photobiol* 1982;35(1):129–132.

Dombrowski H. Bacteria from paleozoic salt deposits. *Ann N Y Acad Sci* 1963;108:453–460.

Dornmayr-Pfaffenhuemer M, Legat A, Schwimbersky K, *et al.* Responses of haloarchaea to simulated microgravity. *Astrobiology* 2011;11(3):199–205.

Doyle L, Ramsay G, Doyle JG. Superflares and variability in solar-type stars with TESS in the Southern hemisphere. *MNRAS* 2020;494:3596–3610.

Fendrihan S, Bérces A, Lammer H, *et al.* Investigating the effects of simulated Martian ultraviolet radiation on *Halococcus dombrowskii* and other extremely halophilic archaeabacteria. *Astrobiology* 2009;9(1):104–112.

Fendrihan S, Dornmayr-Pfaffenhuemer M, Gerbl FW, *et al.* Spherical particles of halophilic archaea correlate with exposure to low water activity—Implications for microbial survival in fluid inclusions of ancient halite. *Geobiology* 2012;10:424–433.

Gladman B. Destination: Earth. Martian meteorite delivery. *Icarus* 1997;130(2):228–246.

Godin PJ, Stone H, Bahrami R, *et al.* UV attenuation by Martian brines. *Can J Phys* 2020;98(6):567–570.

Gooding JL, Wentworth SJ, Zolensky ME. Aqueous alteration of the Nakhla meteorite. *Meteoritics* 1991;26:135–143.

Grant WD, Gemmell RT, McGenity TJ. Halobacteria: The evidence for longevity. *Extremophiles* 1998;2:279–287.

Güdel M. The sun in time: Activity and environment. *Living Rev Sol Phys* 2007;4(3):137.

Güdel M, Audard M, Reale F, *et al.* Flares from small to large: X-ray spectroscopy of Proxima Centauri with XMM-Newton. *A&A* 2004;416:713–732.

Hassenkam T, Andersson MP, Dalby KN, *et al.* Elements of Eoarchean life trapped in mineral inclusions. *Nature* 2017; 548:78–81.

Hiatt CW. Kinetics of the inactivation of viruses. *Bacteriol Rev* 1964;28(2):150–163.

Hidea K, Hayakawa Y, Ito A, *et al.* Wavelength dependence of the formation of single-strand breaks and base changes in DNA by the ultraviolet radiation above 150nm. *Photochem Photobiol* 1986;44(3):379–383.

Horan DM. Kreplin RW. Simultaneous measurements of EUV and soft X-ray solar flare emission. *Sol Phys* 1981;74:265–272.

Horneck G. Survival of microorganisms in space: A review. *Adv Space Res* 1981;1(14):39–48.

Horneck G. Exobiological experiments in Earth orbit. *Adv Space Res* 1998;22(3):317–326.

Horneck G. Astrobiology studies of microbes in simulated interplanetary space. In: *Laboratory Astrophysics and Space Research* (Ehrenfreund P, Krafft C, Kochan H, Pirronello V, eds) Springer: Dordrecht; 1999; pp. 667–685.

Horneck G, Bücker H. *Increased Radiosensitivity of Microorganisms by Vacuum Treatment. International Atomic Energy Agency (IAEA): IAEA.* International Atomic Energy Agency, Vienna (Austria); Food and Agriculture Organization of the United Nations, Rome (Italy); ISBN 92-0-110081-7; September 1981; pp. 95–105; International Symposium on Combination Processes in Food Irradiation, Colombo (Sri Lanka); November 24–28, 1980.

Horneck G, Bücker H, Reitz G. Long-term survival of bacterial spores in space. *Adv Space Res* 1994;14(10):41–45.

Horneck G, Mileikowsky C, Melosh HJ, *et al.* Viable transfer of microorganisms in the Solar System and beyond. In: *Astrobiology* (Horneck G, Baumstark-Khan C. eds.) Springer: Berlin, Heidelberg; 2002; pp. 57–76.

Horneck G, Klaus DM, Mancinelli RL. Space microbiology. *Microbiol Mol Biol Rev* 2010;74(1):121–56.

Im S, Joe M, Kim D, *et al.* Transcriptome analysis of salt-stressed *Deinococcus radiodurans* and characterization of salt-sensitive mutants. *Res Microbiol* 2013;164(9):923–932.

Ito T. Vacuum ultraviolet photobiology with synchrotron radiation. In: *Synchrotron Radiation in Structural Biology* (Sweet RM, Woodhead AD, eds.) Basic Life Sciences, Vol. 51. Springer: Boston, MA; 1989; pp. 221–241.

Karmakar S, Pandey JC, Airapetian VS, *et al.* X-ray superflares on CC Eri. *ApJ* 2017;840:102.

Kauri T, Wallace R, Kushner DJ. Nutrition of the halophilic archaebacterium, *Haloferax volcanii. Syst Appl Microbiol* 1990;13(1):14–18.

Kawaguchi Y, Shin-ichi Y, Hirofumi H, *et al.* Investigation of the interplanetary transfer of microbes in the Tanpopo mission at the exposed facility of the international space station. *Astrobiology* 2016;16(5):363–376.

Koike J, Oshima T, Koike KA, *et al.* Survival rates of some terrestrial microorganisms under simulated space conditions. *Adv Space Res* 1992;12(4):271–274.

Lea D. Actions of radiations on living cells. 1946; Cambridge University Press. London.

Leuko S, Weidler G, Rittmann S, *et al.* LIVE/DEAD Kit: A powerful tool to detect haloarchaeal survival (and life?) in unknown environmental samples. In *Third European Workshop on Exo-Astrobiology*, Madrid, 2004;545:231–232.

Lowenstein TK, Schubert BA, Timofeeff MN. Microbial communities in fluid inclusions and long-term survival in halite. *GSA Today* 2011;21(1):4–9.

Maehara H, Shibayama T, Notsu S, *et al.* Superflares on solar-type stars. *Nature* 2012;485:478–481.

Mancinelli RL, Fahlen TF, Landheim R, *et al.* Brines and evaporites: Analogs for Martian life. *Adv Space Res* 2004;33(8):1244–1246.

Mancinelli RL, White MR, Rothschild LJ. Biopan-survival I: Exposure of the osmophiles *Synechococcus* sp. (Nageli) and *Haloarcula* sp. to the space environment. *Adv Space Res* 1998;22(3):327–334.

Martin O, Peñate L, Cárdenas R, *et al.* The photobiological regime in the very early Earth and the emergence of life. In: *Genesis—In the Beginning* (Seckbach J. ed.) Springer: Dordrecht; 2012; pp. 145–155.

Matsunaga T, Hieda K, Nikaido O. Wavelength dependent formation of thymine dimers and (6-4) photoproducts in DNA by monochromatic ultraviolet light ranging from 150 to 365 nm. *Photochem Photobiol* 1991;54(3):403–410.

McGenity TJ, Gemmell RT, Grant WD, *et al.* Origins of halophilic microorganisms in ancient salt deposits. *Environ Microbiol* 2000;2:243–250.

Melosh HJ. The rocky road to panspermia. *Nature* 1988;332:687–688.

Mileikowsky C, Cucinotta FA, Wilson JW, *et al.* Natural transfer of viable microbes in space: 1. From Mars to Earth and Earth to Mars. *Icarus* 2000;145(2):391–427.

Moissl-Eichinger C, Cockell C, Rettberg P. Venturing into new realms? Microorganisms in space. *FEMS Microbiol Rev* 2016;40(5):722–737.

Mojzsis SJ, Arrhenius G, McKeegan KD, *et al.* Evidence for life on Earth before 3,800 million years ago. *Nature* 1996;384(6604):55–59.

Mojzsis SJ, Brasser R, Kelly NM, *et al.* Onset of giant planet migration before 4480 million years ago. *ApJ* 2019;881(1):44.

Moore JM, Bullock MA, Newsom H, *et al.* Laboratory simulations of Mars evaporite geochemistry. *J Geophys Res* 2010;115:E06009.

Nicholson WL. Ancient micronauts: Interplanetary transport of microbes by cosmic impacts. *Trends Microbiol* 2009;17(6):243–250.

Nicholson WL, Schuerger AC, Setlow P. The solar UV environment and bacterial spore UV resistance: Considerations for panspermia and planetary protection. *Mutat Res* 2005;571:249–264.

Nisbet EG, Sleep NH. The habitat and nature of early life. *Nature* 2001;409(6823):1083–1091.

Noffke N, Christian D, Wacey D, *et al.* Microbially induced sedimentary structures recording an ancient ecosystem in the ca. 3.48 billion-year-old Dresser Formation, Pilbara, Western Australia. *Astrobiology* 2013;13(12):1103–1124.

Norton CF, McGenity TJ, Grant WD. Archaeal halophiles (halobacteria) from two British salt mines. *J Gen Microbiol* 1993;139:1077–1081.

Okamoto TJ, Sakurai T. Super-strong magnetic field in sunspots. *Astrophys J Lett* 2018;852:L16.

Olsson-Francis K, Cockell CS. Experimental methods for studying microbial survival in extraterrestrial environments. *J Microbiol Methods* 2010;80(1):1–13.

Oren A. The Family Halobacteriaceae. In: *The Prokaryotes. A Handbook on the Biology of Bacteria: Ecophysiology and Biochemistry, 4th ed.* (Rosenberg E, DeLong EF, Thompson F, Lory S, Stackebrandt E. eds.) Springer: New York, NY, USA; 2014; pp. 41–121.

Paulino-Lima IG, Pilling S, Janot-Pacheco E, *et al.* Laboratory simulation of interplanetary ultraviolet radiation (broad spectrum) and its effects on *Deinococcus radiodurans. Planet Space Sci* 2010;58(10):1180–1187.

Pennell KG, Aronson AI, Blatchley III ER. Phenotypic persistence and external shielding ultraviolet radiation inactivation kinetic model. *J Appl Microbiol* 2008;104(4):1192–1202.

Portner DM, Spiner DR, Hoffman RK, *et al.* Effect of ultrahigh vacuum on viability of microorganisms. *Science* 1961;134(3495):2047.

Pye JP, Rosen S, Fyfe D, *et al.* A survey of stellar X-ray flares from the XMM-Newton serendipitous source catalogue: HIPPARCOS-Tycho cool stars. *A&A* 2015;581:A28.

Rabbow E, Rettberg P, Barczyk S, *et al.* EXPOSE-E: An ESA astrobiology mission 1.5 years in space. *Astrobiology* 2012; 12(5):374–386.

Reiser R, Tasch P. Investigation of the viability of osmophile bacteria of great geological age. *Trans Kans Acad Sci* 1960; 63:31–34.

Ribas I, Guinan EF, Güdel M, *et al.* Evolution of the solar activity over time and effects on planetary atmospheres. I. High-energy irradiances (1–1700 Å). *ApJ* 2005;622:680–694.

Ribas I, Porto de Mello GF, Ferreira LD, *et al.* Evolution of the solar activity over time and effects on planetary atmospheres. II. κ¹Ceti, an analog of the Sun when life arose on Earth. *ApJ* 2010;714:384–395.

Richter H. On the darwinian doctrine. [Zur Darwinschen Lehre.] *Schmidts Jahrbuch Ges. Med* 1865;126:243–249.

Robinson CR, Bopp BW. A ''Helium Flare'' on the Active G5 Dwarf Kappa Ceti. In: *Cool Stars, Stellar Systems and the Sun* (Linsky JL, Stencel RE, eds) Vol. 291 Springer: Berlin, Heidelberg; 1987; pp. 509–511.

Roedder E. Volume 12: Fluid inclusions. *Rev Mineral* 1984;12: 644.

Rosing MT. ¹³C-depleted carbon microparticles in >3700-Ma sea-floor sedimentary rocks from West Greenland. *Science* 1999;283:674–676.

Rosing MT, Rose NM, Bridgwater D, *et al.* Earliest part of Earth's stratigraphic record: A reappraisal of the >3.7 GaIsua (Greenland) supracrustal sequence. *Geology* 1996;24:43–46.

Saenger W. Water and Nucleic Acids. In: *Principles of Nucleic Acid Structure* (Cantor CR, eds.) Springer: New York, NY, USA; 1984; pp. 368–384.

Saffary R, Nandakumar R, Spencer D, *et al.* Microbial survival of space vacuum and extreme ultraviolet irradiation: Strain isolation and analysis during a rocket flight. *FEMS Microbiol Lett* 2002;215(1):163–168.

Sankaranarayanan K, Lowenstein TK, Timofeeff MN, *et al.* Characterization of ancient DNA supports long-term survival of Haloarchaea. *Astrobiology* 2014;14(7):553–560.

Sanz-Forcada J, Brickhouse NS, Dupree AK. The structure of stellar coronae in active binary systems. *ApJS* 2003;145:147–179.

Sanz-Forcada J, Micela G. The EUVE point of view of AD Leo. *A&A* 2002;394:653–661.

Sanz-Forcada J, Micela G, Ribas I, *et al.* Estimation of the XUV radiation onto close planets and their evaporation. *A&A* 2011; 532:A6.

Sanz-Forcada J, Ribas I. High-energy irradiances of Sun-like stars. In: *Pathways Towards Habitable Planets*. Proceedings of a Conference held at Bern, Switzerland, July 13–17, 2015.

Satterfield CL, Lowenstein TK, Vreeland RH, *et al.* New evidence for 250 Ma age of halotolerant bacterium from a Permian salt crystal. *Geology* 2005;33(4):265–268.

Schaefer BE, King JR, Deliyannis CP. Superflares on ordinary solar-type stars. *ApJ* 2000;529:1026–1030.

Schidlowski M. Clay minerals and the origin of life. *Earth Sci Rev* 1988;25(4):326–327.

Schidlowski M. Carbon isotopes as biogeochemical recorders of life over 3.8 Ga of Earth history: Evolution of a concept. *Precambr Res* 2001;106:117–134.

Schindelin J, Arganda-Carreras I, Frise E, *et al.* Fiji: An opensource platform for biological-image analysis. *Nat Methods* 2012;9(7):676–682.

Schubert BA, Lowenstein TK, Timofeeff MN, *et al.* Halophilic archaea cultured from ancient halite, Death Valley, California. *Environ Microbiol* 2010;12:440–454.

Schubert BA, Lowenstein TK, Timofeeff MN, *et al.* How do prokaryotes survive in fluid inclusions in halite for 30,000 years? *Geology* 2009b;37:1059–1062.

Schubert BA, Lowenstein TK, Timofeeff MN, *et al.* *Environ Microbiol* 2010;12:440–454.

Schwarzer S, Rodriguez-Franco M, Oksanen HM, *et al.* Growth phase dependent cell shape of *Haloarcula*. *Microorganisms* 2021;9(2):231.

Shibata K, Isobe H, Hillier A. Can superflares occur on our Sun? *Publ Astron Soc Jpn* 2013;65:49.

Slade D, Radman M. Oxidative stress resistance in *Deinococcus radiodurans*. *Microbiol Mol Biol Rev* 2011;75(1):133–191.

Sontag W, Dertinger H. Energy requirements for damaging DNA molecules: III. The mechanisms of inactivation of bacteriophage ΦX 174 DNA by vacuum ultraviolet radiation. *Int J Radiat Biol Relat Stud Phys Chem Med* 1975;27(6):543–552.

Stan-Lotter H, Fendrihan S. Halophilic archaea: Life with desiccation, radiation and oligotrophy over geological times. *Life* 2015;5(3):1487–1496.

Stan-Lotter H, Radax C, McGenity TJ, *et al.* From Intraterrestrials to Extraterrestrials—Viable Haloarchaea in Ancient Salt Deposits. In: *Halophilic Microorganisms* (Ventosa A, ed) Springer Verlag: Berlin, Heidelberg, New York, NY, USA; 2004; pp. 89–102.

Stocks SM. Mechanism and use of the commercially available viability stain BacLight. *Cytometry A* 2004;61(2):189–195.

Stöffler D, Horneck G, Ott S, *et al.* Experimental evidence for the potential impact ejection of viable microorganisms from Mars and Mars-like planets. *Icarus* 2007;186:585–588.

Swarts SG, Sevilla MD, Becker D, *et al.* Radiation-induced DNA damage as a function of hydration: I. Release of unaltered bases. *Radiat Res* 1992;129(3):333–344.

Takakura K, Ishikawa M, Hidea K, *et al.* Single-strand breaks in supercoiled DNA induced by vacuum-UV radiation in aqueous solution. *Photochem Photobiol* 1986;44(3):397–400.

Tasch P. Fossil Content of Salt and Association Evaporates. In: *Symposium on Salt.* Cleveland: Northern Ohio Geological Society; 1963;1; pp. 96–102.

Tashiro T, Ishida A, Hori M, *et al.* Early trace of life from 3.95 Ga sedimentary rocks in Labrador, Canada. *Nature* 2017; 549(7673):516–518.

Thomson W. The British Association Meeting at Edinburgh: Inaugural address of Sir William Thomson, LL.D., F.R.S., President. *Nature* 1871;4:261–278.

Tu Z-L, Yang M, Zhang ZJ, *et al.* Superflares on solar-type stars from the first-year observation of TESS. *ApJ* 2020;890:46.

Vreeland RH, Jones J, Monson A, *et al.* Isolation of live Cretaceous (121–112 million years old) halophilic Archaea from primary salt crystals. *Geomicrobiol J* 2007;24:275–

Wehner J, Horneck G. Effects of vacuum UV and UVC radiation on dry *E. coli* plasmid pUC19 I. Inactivation, *lacZ⁻* mutation induction and strand breaks. *J Photochem Photobiol B Biol* 1995;28(1):77–85.

Wirths A, Jung H. Single-strand breaks induced in DNA by vacuum-ultraviolet radiation. *Photochem Photobiol* 1972; 15(4):325–330.

Yamada H, Hieda K. Wavelength dependence (150–290 nm) of the formation of the cyclobutane dimer and the (6–4) photoproduct of thymine. *Photochem Photobiol* 1992;55(4):541–548.

Yamagishi A, Kawaguchi Y, Hashimoto H, *et al.* Environmental data and survival data of *Deinococcus aetherius* from the exposure facility of the Japan experimental module of the international space station obtained by the Tanpopo mission. *Astrobiology* 2018;18(11):1369–1374.

Zolensky ME, Bodnar RJ, Gibson Jr EK, *et al.* Asteroidal water within fluid inclusion-bearing halite in an H5 chondrite, Monahans (1998). *Science* 1999;285(5432):1377–9.

Address correspondence to:
*Ximena C. Abrevaya*
*Instituto de Astronomía y Física del Espacio (IAFE)*
*UBA–CONICET*
*Pabellón IAFE*
*Ciudad Universitaria*
*C1428EGA Ciudad Autónoma de Buenos Aires*
*Argentina*

*E-mail:* abrevaya@iafe.uba.ar

| Abbreviations Used | |
| --- | --- |
| CFU = | colony-forming units |
| CNPEM = | Brazilian Center for Research in Energy and Materials |
| CS = | cell suspensions |
| DS = | deep survey |
| EUV = | extreme ultraviolet |
| EUVE = | Extreme Ultraviolet Explorer |
| LHB = | Late Heavy Bombardment |
| LNLS = | Brazilian Synchrotron Light Laboratory |
| LOD = | limit of detection |
| PI = | propidium iodide |
| QE = | quantum efficiency |
| ROS = | reactive oxygen species |
| SD = | standard deviation |
| TGM = | toroidal grating monochromator |
| UV = | ultraviolet |
| VUV = | vacuum-ultraviolet |
| XUV = | X-ray and extreme ultraviolet |